\documentclass[useAMS,usenatbib]{mn2e}
\usepackage{epsfig}
\usepackage{amsmath}
\usepackage[firstpage]{draftwatermark}

   \usepackage{epstopdf}
\newcommand{\btxt}[1]{{\bf #1}}

\title[Stellar Mass and Star Formation in BGGs]{Brightest Group Galaxies: Stellar Mass and Star Formation Rate (paper I)}
\author[Gozaliasl et al.]{Ghassem Gozaliasl$^{1}$\thanks{E-mail: ghassem.gozaliasl@helsinki.fi},
Alexis Finoguenov$^{1}$, Habib G. Khosroshahi$^{2}$, \newauthor Mohammad Mirkazemi$^{3}$, Ghazaleh Erfanianfar$^{3}$, Masayuki Tanaka$^4$\\
$^1$Department of Physics, University of Helsinki, P. O. Box 64, FI-00014, Helsinki, Finland\\
$^2$School of Astronomy, Institute for Research in Fundamental Sciences (IPM), P. O. Box 19395-5531, Tehran, Iran\\
$^3$Excellence Cluster Universe, Boltzmannstr. 2, 85748 Garching bei München, Germany\\
$^4$ National Astronomical Observatory of Japan, 2-21-1 Osawa, Mitaka, Tokyo 181-8588, Japan}

\begin{document}

\maketitle


\label{firstpage}
\begin{abstract}

We study the distribution and evolution of the stellar mass and the star formation rate (SFR) of the brightest group galaxies (BGGs)  over $ 0.04<z<1.3 $ using a large sample of $ 407 $ X-ray galaxy groups selected from the COSMOS, AEGIS, and XMM-LSS fields. We compare our results with predictions from the semi-analytic models based on the Millennium simulation. In contrast to  model predictions, we find that, as the universe evolves, the stellar mass distribution evolves towards a normal distribution. This distribution tends to skew to low mass BGGs at all redshifts implying the presence of a star forming population of the BGGs with $ M_S\sim10^{10.5} M_{\odot} $ which results in the shape of the stellar mass distribution  deviating from a normal distribution. In agreement with the models and previous studies, we find that the mean stellar mass of BGGs grows with time by a factor of $\sim2$ between $z=1.3$ to $z=0.1$, however, the significant growth occurs above $ z=0.4$. The BGGs  are not entirely a dormant population of galaxies, as low mass BGGs in low mass halos are more active in forming stars than the BGGs in more massive halos, over the same redshift range. We find that the average SFR of the BGGs evolves steeply with redshift and fraction of the passive BGGs increases as a function of increasing  stellar mass and halo mass. Finally, we show that the specific SFR of the BGGs within halos with $ M_{200} \leq 10^{13.4} M_{\odot} $ decreases with increasing halo mass at  $ z<0.4 $.
\end{abstract}
\begin{keywords}
galaxies: clusters: general � galaxies: elliptical and lenticular, cD � galaxies:
haloes � intergalactic medium � X-ray: galaxies � X-rays: galaxies: clusters.\end{keywords}

\section{Introduction} \label{Introduction}
Brightest groups/cluster galaxies (here after BGGs) are generally located at the core of the hosting halos and close to the centre of the extended X-ray emitting hot intragroup/cluster medium. Their privileged location within the group/cluster and their stellar luminosity makes them ideal targets for constraining cosmological models and studying the assembly histories of massive galaxies \citep{sandage72,GunnOke75,sandage76,HoesselSchnieder85,
bhavsar85,oegerle91,postman95,bernstein01,bernardi07,vonderlinden07,liu09,deLucia07,stott10}.
 
BGGs are generally early type galaxies with no significant ongoing star formation. They differ from other massive galaxies in surface brightness profiles and scaling relations \citep[e.g.][]{bernardi07,vonderlinden07,Liu08,Graham96}. These findings suggest that formation of the BGGs could be different from that of other normal elliptical galaxies
\citep[e.g.][]{stott08,Collins98,stott10,Burke00,Wen11,Mendez12,Shen14}. 

Understanding the stellar mass assembly of galaxies, particularly BGGs,  is highly important for galaxy formation and evolution scenarios and thus has been  the focus of a number of studies based on the hierarchical assembly of the dark matter halos in the $ \Lambda $ cold dark matter $ (\Lambda CDM) $ cosmology \citep[e.g.][]{White78,Voit05}; halo abundance matching simulations \cite[e.g.][]{Moster13}; the semi-analytic model (SAM) of galaxy formation \cite[e.g.][]{Bower06,Croton06,deLucia07,Guo11,Cousin15,Tonini12}; and the observational studies as well
\cite[e.g.][]{stott10,Lin13,Oliva14}. 

Previous studies indicate that the evolutionary history of galaxies in terms of the mass growth can be divided into two main epochs. A very early epoch, above $ z\sim2 $, in which the star formation rate (SFR) is peaked and the majority of stars  ($ \sim80\% $) in the BGGs form due to the rapid cooling from their hot and dense halos \cite[e.g.][]{deLucia07,Hopkin06}. A later epoch, SFR declines possibly because the cold stream is suppressed by shock heating \citep{Dekel09}. The stellar and AGN feedback may have played a role as well. Furthermore, strong ram-pressure stripping of the interstellar medium and the tidal fields of groups and clusters contributed to the quenching. By this stage, the BGGs have considerably grown in mass and they generally follow a passive evolution \citep[e.g.][]{Liu10,conroy07,stott10,Croton06}. 

Using near-infrared luminosity as a proxy for stellar mass, some studies determined the growth of the stellar mass in BGGs, claiming that these systems exhibit little/no changes in mass below $ z\sim1$ \citep{Collins98,Whiley08,stott10,Collins09}. In contrast, some recent studies have  argued that the stellar mass of the BGGs have grown due to galaxy mergers by a factor of $ \sim2$ between $ z\sim1 $ and $z\sim0.2$ \citep{deLucia07,Lidman12,Bai14,Lin13}. These contradicting results demonstrate that the details of the stellar mass assembly of BGGs still remain elusive. The main purpose of this paper is to take advantage of a large sample of X-ray detected groups and their BGGs to study the stellar mass  evolution over the last  $\sim9$ billions years of the age of the universe.  

Semi-analytic models have become a standard tool to help us interpret observations and place constraints on the formation and evolution of galaxies. Using a semi analytic model based on the Millennium simulation \citep{Springel05}, \cite{deLucia07} traced back the formation history of the central cluster galaxies within dark matter halos and pictured their full merger trees. They found that significant fraction of stars (50\% and 80\%) in the central luminous galaxy form  at high redshifts (above  $z \sim 3$ and $z \sim 2$) and half of their final mass is carried by a single galaxy below $ z\sim 0.5$. Since then, they grow through major mergers and a significant number of the minor mergers. Furthermore, \cite{Lin13}, using a sample of the BGGs selected from the {\it Spitzer} IRAC Shallow Cluster Survey with halo mass range of $ (2.4-4.5)\times 10^{14} $, found that these galaxies have grown by a factor of 2.3 between $ z=1.5 $ and $ z=0.5 $. While the predictions from  \cite{Guo11} SAM agrees well with these observations, below $ z<0.5 $  the model fails to reproduce the observed growth. 

Moreover, \cite{Liu10} studied the stellar mass of central and satellite galaxies selected from the  galaxy group catalogs  in the Sloan Digital Sky Survey (SDSS) and  the  three SAMs, and found that all  models  fail to  reproduce  the  sharp  decrease  in the stellar  mass  with  decreasing  halo  mass  at  the  low  mass end while models also over-predict the number of satellites by roughly a factor of 2. Furthermore, we have shown that SAMs fail to match  the observed  evolution of the relation between the magnitude gap between the first the second brightest group galaxies, and their intrinsic luminosity \citep{Gozaliasl14,Smith10}. Thus, the SAMs still requires further improvement and more sophisticated treatment of the physical processes, which is often performed via tuning a large number of parameters \citep{benson10}. However, they still serve as a reasonable guide, we compare three SAMs based on the Millennium simulation presented in \citet[][hereafter B06]{Bower06}, \citet[][hereafter DLB07]{deLucia07}, and \citet[][hereafter G11]{Guo11} with the observations.

In this paper, we analyse  a large sample of the X-ray galaxy groups and clusters selected from the XMM-LSS \citep{Gozaliasl14}, COSMOS \citep{Finoguenov07,George11} and AEGIS \citep{erfanianfar13} fields in a wide redshift range $ 0.04<z<1.3 $. Most of the previous studies have focused on the evolution of the brightest galaxies within massive clusters with masses above $ M_{200}\sim10^{14} M_{\odot}$. Our sample covers a lower mass range between $M_{200}\sim10^{12.8} $ and $ 10^{14} M_{\odot}$. This allows us to study the galaxy assembly in more common environments in the universes.  

This paper is structured as follows. We define data and describe how we infer galaxy stellar mass in section 2. In section 3 we present the distribution and evolution of the stellar mass of the BGGs. Section 4, presents an analysis of the distribution of the star formation rate in the BGGs and  its dependence on  stellar/halo mass. Section 5 is a summary of  our results.
 
 Unless stated otherwise, we adopt a cosmological model, with $(\Omega_{\Lambda}, \Omega_{M}, h) = (0.75, 0.25, 0.71$), where the Hubble constant is characterized as 100 h km s$^{-1}$ Mpc$^{-1}$ and quote uncertainties on 68\% confidence level. 
 \begin{figure*}
   \includegraphics[width=6cm]{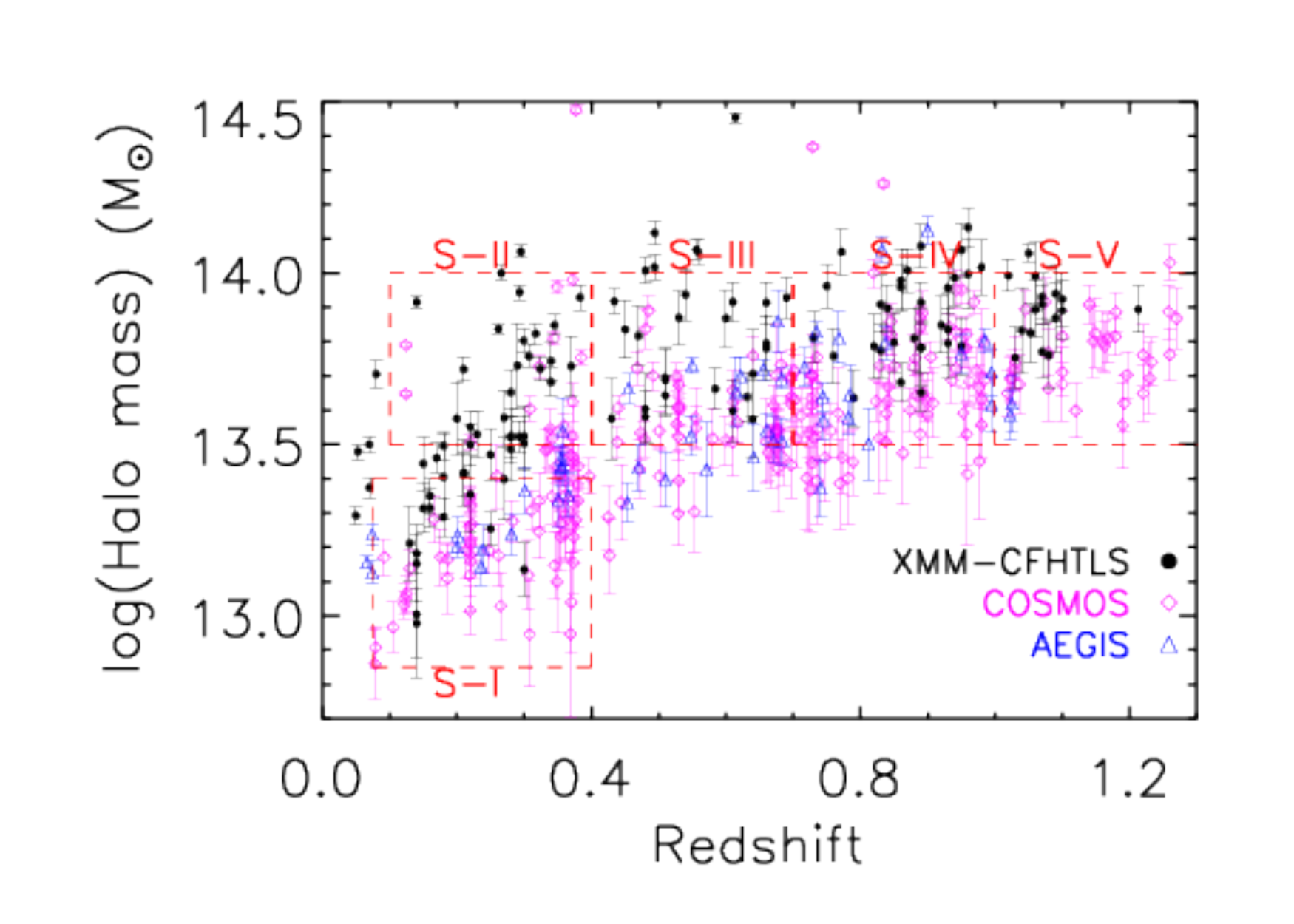}
  \includegraphics[width=5cm]{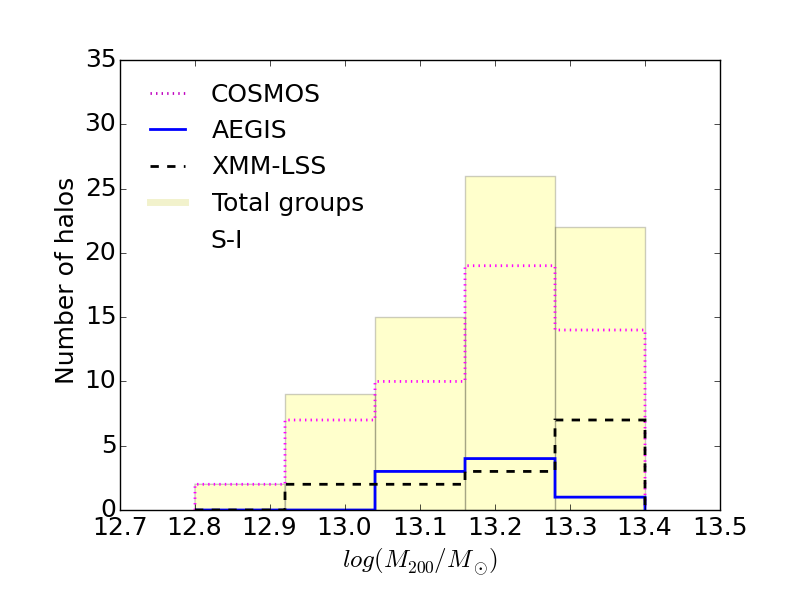}
  \includegraphics[width=5cm]{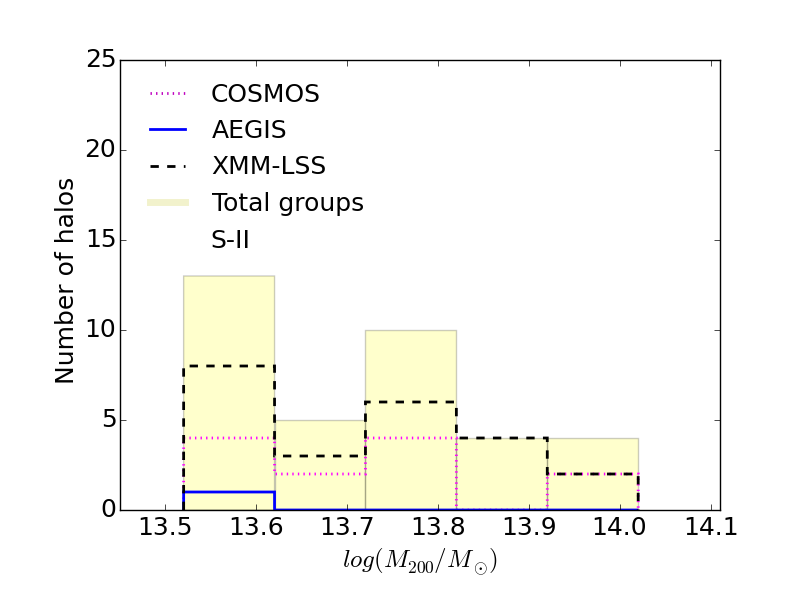}
  \includegraphics[width=5cm]{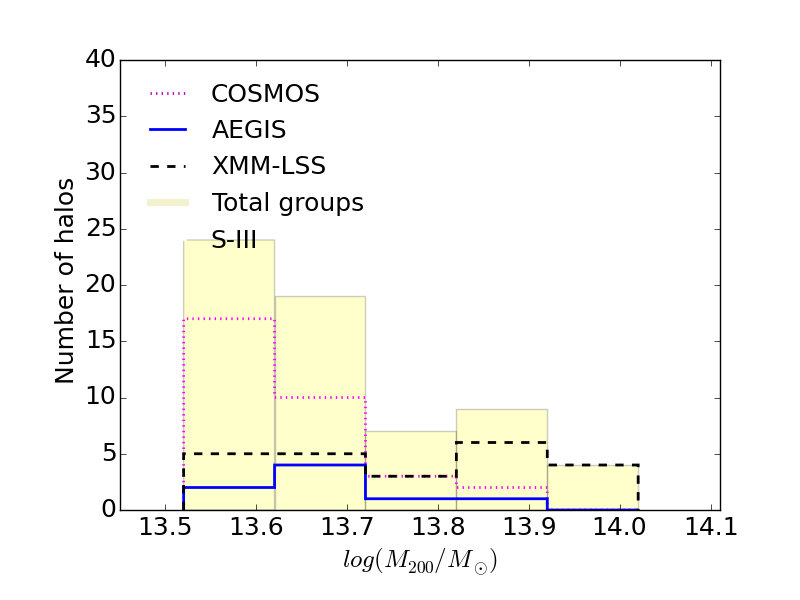}
  \includegraphics[width=5cm]{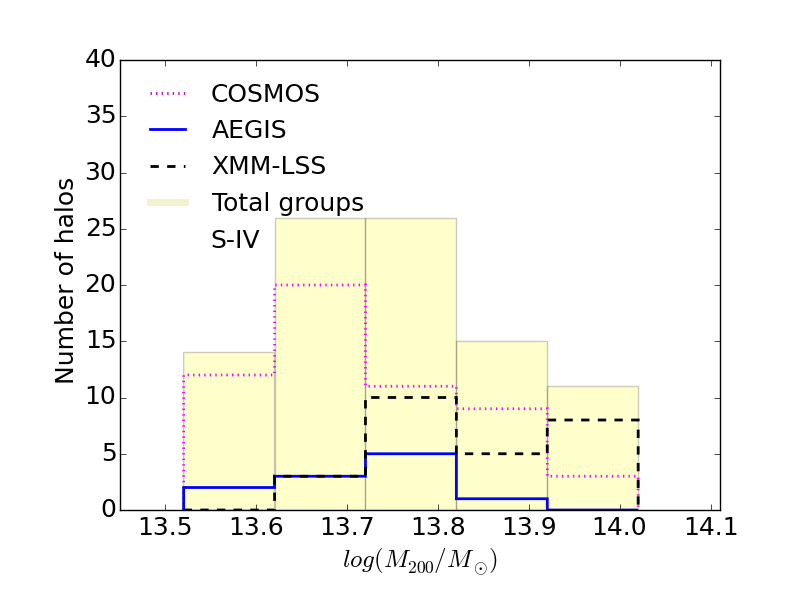}
  \includegraphics[width=5cm]{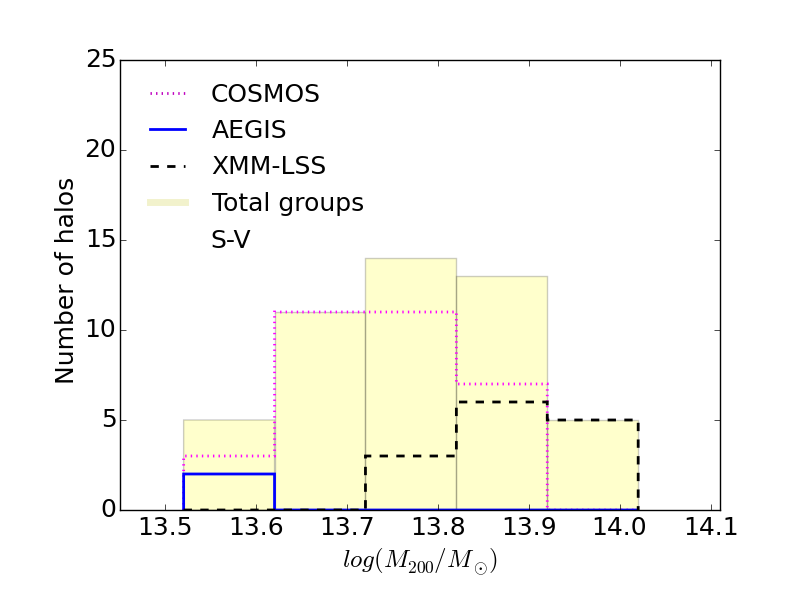}

  \caption[]{\textit{(Left upper panel)} $ M_{200} $ as a function of  the redshift for the X-ray galaxy groups selected from the COSMOS (open
    magenta diamonds), XMM--LSS (filled black circles ), and AEGIS (open blue triangles) fields. The dashed red boxes
    present our five defined subsamples as described in \S \ref{subsample}. \textit{(All panels except the left upper panel)}  The  halo mass distribution  of the total number of  galaxy groups within each subsample (filled yellow histogram)  and  those  selected from the  COSMOS (dashed magenta histogram), XMM--LSS (solid black histogram ), and AEGIS (dotted blue histogram) fields.   }
  \label{mass}
  \end{figure*}
\section{Group sample}\label{sample}
\subsection{Subsample definition}
\label{subsample}
We use the combined data of the X-ray galaxy groups and clusters from 3 surveys, XMM--LSS, COSMOS, and AEGIS,  which have been presented in details in \cite{Gozaliasl14,Finoguenov07,George11,erfanianfar13}. These catalogs include 456 galaxy groups/clusters with halo masses ranging from $M_{200}\sim5\times10^{12} $ to $10^{14.5}(M_{\odot}) $  over a redshift range of  $0.04<z<1.9$.

 Over 70 per cent of all galaxy groups in our data have  spectroscopic redshifts.
For the purpose of our study and to ensure a high accuracy photometric redshift measurement, we limit  the sample to the redshift range of $ 0.04\le z \le 1.3 $ and halo mass range of $ M_{200}\simeq7.25\times10^{12} $ to $ 1.04\times10^{14} (M_{\odot}) $, namely galaxy groups. This selection provides us with a sample of $407$ X-ray groups. The left upper panel of Fig. \ref{mass} illustrates the halo mass ($M_{200}$) as a function of the group redshift. This plane allows us to define five subsamples as
follows:

(S-I)   $0.04 <$ z $< 0.40$ $ \& $ $12.85 < log(\frac{M_{200}}{M_{\odot}})  \le 13.40 $ 
 
(S-II) $0.10 <$ z $\leq 0.4$ $ \& $  $13.50 < log(\frac{M_{200}}{M_{\odot}}) \le 14.02 $
 
(S-III) $0.4 <$ z $\leq 0.70$ $ \& $  $13.50 < log(\frac{M_{200}}{M_{\odot}}) \le 14.02 $
  
(S-IV) $0.70 <$ z $\leq 1.0$ $ \& $  $13.50 < log(\frac{M_{200}}{M_{\odot}}) \le 14.02 $

(S-V) $1.0 <$ z $\leq 1.3$ $ \& $  $13.50 < log(\frac{M_{200}}{M_{\odot}}) \le 14.02 $ \\
          
The defined subsamples (hereafter, S-I to S-V) include 74, 36, 63, 92, and 48 galaxy groups, respectively.  The $
M_{200}-z $ plane enables us to adopt  a similar halo mass range for the last four subsamples (S-II to S-V). The halo mass range  for these subsamples is narrow ($<0.5$ dex). The average uncertainty  associated to the mass estimate of our sample of galaxy groups is about $ 0.17\pm0.1 $ dex. In Fig. \ref{mass} we separately show the halo mass distribution for all subsamples of galaxy groups and those selected from the COSMOS (dashed magenta histogram), XMM--LSS (solid black histogram ), and AEGIS (dotted blue histogram) fields. The halo mass distribution for the total number of the galaxy groups within  S-II to S-V  peaks around  $log(M_{200}/M_\odot)\sim13.7\pm0.2$. For S-II and S-III, the peak of distribution tends to skew lower masses. However, we repeat our analysis several times with considering different halo mass ranges for the last four subsamples and find  that the small bias towards high halo masses at high redshifts falls within errors and dose not affect our results. Thus, this also makes it possible to investigate the evolutionary properties  of the BGGs  over $ 0.1<z<1.3 $.  

In this paper, we compare observations with predictions from three SAMs  (G11, DLB07 and B06)  based on the Millennium simulation. For  detailed information on these  models, we refer reader to \cite{Guo11,deLucia07,Bower06}. We note that  some important properties of these models have been summarized in \cite{Gozaliasl14}. 
We randomly select our subsamples of galaxy groups  from the SAM catalogs according to the redshift and halo mass limits  which  defined  for the sample of the galaxy groups  in observations. 
\subsection{BGG selection}
\begin{figure}
     \includegraphics[width=7cm]{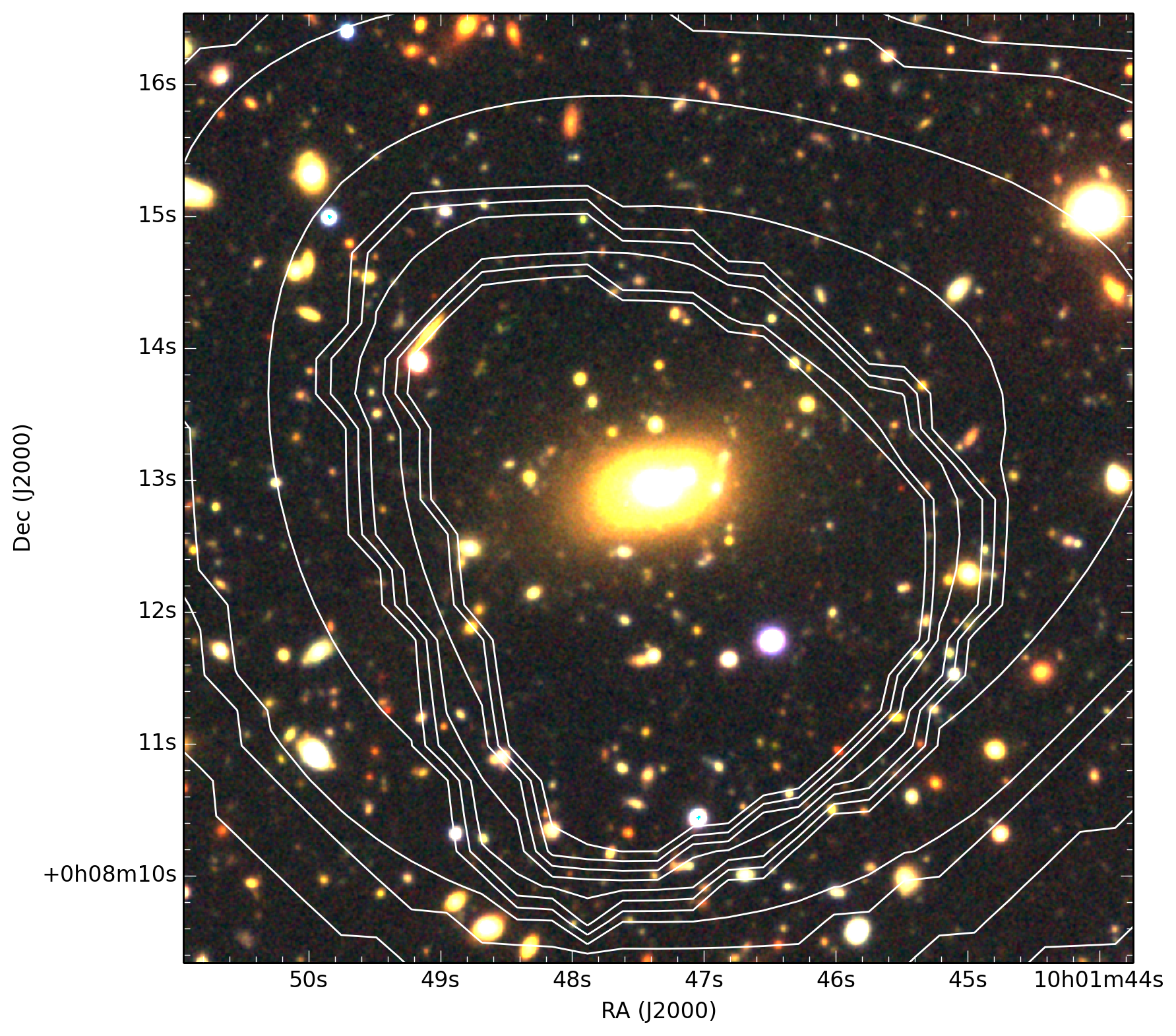}
     \caption[]{ The X-ray emission contours overlaid on the optical RGB image of a galaxy group at $z=0.322$ in the COSMOS
     field. The brightest group galaxy  is located in the X-ray center of the hosting group. }
     \label{rgb}
     \end{figure}

We first identify member galaxies with spectroscopic redshift and then include likely member galaxies with red sequences with a help of multi-band photo-z's. The galaxy photometric redshift catalogs are from \cite{Ilbert13,McCracken12, Capak07} (COSMOS),  \cite {Brimioulle08,Brimioulle13} (XMM-LSS), and \cite{Wuyts11} (AEGIS), and the red sequences finder can be found in \cite{Mirkazemi15}. From the member catalogs, we define the brightest galaxy as BGG. Note that more than 50\% of the BGGs have spectroscopic redshift. Fig. \ref{rgb} shows a sample BGG at z=0.322.  

Furthermore, we visually inspect all the BGGs in the optical RGB image of hosting halos  (e.g. Fig. \ref{rgb}).   
 
  \subsection{Stellar mass of the BGGs} \label{sm_cal}
     
 The stellar mass and star-formation rate are computed using $Le Phare$ code (\citealt{Arnouts02}; \citealt{Ilbert06}). To start with, the redshift of BGG is fixed to the redshift of the hosting group. Then, similar to \cite{Ilbert10}, SED templates of galaxies generated by \cite{Bruzual03}  are fitted to the photometric measurements in $u$, $g$, $r$, $i$, and $z$  bands using a  $\chi^2$ minimisation method.  For consistency  between data of different fields, we  apply the same method as \cite{Ilbert10} applied for the COSMOS sample.

The SED templates are generated assuming 0.02 and 0.008 metallicities, corresponding to 1 Z$_{\odot}$ and 0.4 Z$_{\odot}$, respectively,  and exponentially declining star formation rate  SFR $\sim$ $\propto  e^{-t/\tau}$ where $t$ is the age of a galaxy and $\tau$ have nine values  between 0.1 to 30 Gyr. The \cite{Calzetti00} extinction law is applied to the SEDs with six $E(B-V)$ values of 0.0, 0.1, 0.2, 0.3, 0.4, and 0.5. Similar to \cite{Ilbert10,Fontana06,Pozzetti07}, we force the prior $E(B-V) <$ 0.15 if age/$\tau$ $>$ 4 (a high extinction value is only allowed for galaxies with a high SFR).
   
In order to estimate the uncertainties in computing the stellar mass, we take into account  the  following sources of errors: 1) uncertainties in redshift determination, 2) lack of near-infrared bands; 3) photometric  errors, and 4) an intrinsic error caused by SED fitting method. For the first source of error, similar to  \cite{Ilbert10}, we compute the stellar mass for a galaxy sample with spectroscopic redshifts. Then we repeat the procedure by fixing their redshift to the one obtained photometrically. To measure the error in stellar mass induced by redshift uncertainty, we compare the stellar mass estimated using  spectroscopic versus photometric redshifts. Since the photometric redshift error depends on photometric accuracy (generally brighter galaxies have more precise photometric redshift), the error induced by this uncertainty is also a function of galaxy luminosity. The error due to the lack of the near-infrared bands and photometric errors is derived by comparing the results of the stellar mass in the CFHTLS deep fields (in overlap with wide fields and with additional $J$, $H$, and $K$ bands). This uncertainty is characterised as a function of the magnitude and redshift of galaxies. Similar to \cite{Giodini12}, we assumed 0.14 dex  error induced by SED fitting method. In order to derive the total errors in the stellar mass estimation for XMM-LSS galaxies, we assume no correlation between aforementioned errors and thus add them in quadrature. Details of error analysis are presented in Mirkazemi et al. (in preparation).   
   
For the BGGs in the COSMOS field, we use the information on the properties of galaxies provided by \cite{Ilbert13,Capak07,McCracken12}. We note that the stellar mass and star formation rate of galaxies in the AEGIS field have been estimated using FAST code \citep{Kriek09} (which has been taken from \cite{Wuyts11}). 
 \begin{table*}
\caption[]{ \btxt{The best-fit Gaussian distribution's parameters.}}
   \begin{tabular}{llll} 
   \hline\hline  
    Sample ID &  Peak value& Peak centroid ($ log (M_{cen}/M_{\odot}) $) & Gaussian width ($ \sigma $) \\
    \hline
\textbf{S-I:}\\
     Obs &  $0.24\pm0.02 $ & $ 10.9\pm 0.01$ & $0.35\pm0.01$\\
	 G11& $ 0.19 \pm 0.01$ & $ 10.94\pm 0.01 $ &$0.19\pm0.01$\\
	 DLB07&$ 0.21 \pm 0.01 $ & $ 11.01\pm 0.01 $ & $0.18\pm0.01$\\
     B06& $ 0.11 \pm 0.01 $ & $ 10.71\pm 0.01  $&$0.32\pm0.01$\\
\hline
\textbf{S-II:}\\
Obs &  $0.18\pm0.03$ & $ 11.10\pm 0.03$ & $0.33\pm0.03$\\
	 G11& $ 0.20 \pm 0.01$ & $ 11.17\pm 0.01 $ &$0.18\pm0.01$\\
	 DLB07&$ 0.22 \pm 0.01 $ & $ 11.25\pm 0.01 $ & $0.16\pm0.01$\\
     B06& $ 0.12\pm 0.01 $ & $ 10.97\pm 0.01  $&$0.260\pm0.01$\\
\hline
\textbf{S-III:}\\
Obs &  $0.20\pm0.02$ & $ 11.25\pm 0.03$ & $0.27\pm0.03$\\
	 G11& $ 0.22 \pm 0.01$ & $ 11.14\pm 0.01 $ &$0.17\pm0.01$\\
	 DLB07&$ 0.21 \pm 0.02 $ & $ 11.18\pm 0.01 $ & $0.19\pm0.01$\\
     B06& $ 0.13\pm 0.01 $ & $ 10.98\pm 0.01  $&$0.32\pm0.03$\\
\hline
\textbf{S-IV:}\\
Obs &  $0.26\pm0.02$ & $ 11.15\pm 0.02$ & $0.37\pm0.02$\\
	 G11& $ 0.20 \pm 0.01$ & $ 11.10\pm 0.01 $ &$0.21\pm0.01$\\
	 DLB07&$ 0.20 \pm 0.01 $ & $ 11.14\pm 0.01 $ & $0.21\pm0.01$\\
     B06& $ 0.13\pm 0.01 $ & $ 10.88\pm 0.01  $&$0.30\pm0.01$\\
\hline
\textbf{S-V:}\\
Obs &  $0.25\pm0.03$ & $ 11.02\pm 0.02$ & $0.35\pm0.04$\\
	 G11& $ 0.27 \pm 0.01$ & $ 11.05\pm 0.01 $ &$0.20\pm0.01$\\
	 DLB07&$ 0.31 \pm 0.02 $ & $ 11.09\pm 0.01 $ & $0.19\pm0.01$\\
     B06& $ 0.19\pm 0.01 $ & $ 10.76\pm 0.01  $&$0.27\pm0.01$\\
  \hline
  
 \end{tabular}
 \label{gg}
 \end{table*}
    
 \begin{table*}
\caption[]{The kurtosis parameter of the stellar mass distribution of the BGGs in observations and SAMs, which are
determined with respect to the normal distribution.}
 \begin{tabular}{llllll} 
 \hline\hline  
  Subsample ID &  mean z & observations &   G11&  DLB07&  B06 \\
  \hline
  
  S-I& 0.22  & $ -0.12 \pm 0.53 $ & $ 0.02 \pm 0.15 $ & $ -0.19 \pm 0.15 $ & $ 1.00 \pm 0.08$\\ 
 S-II& 0.25  &$ -0.07  \pm 0.63 $&$ -0.34 \pm 0.22 $&$ -0.05  \pm 0.24 $&$ 0.05  \pm 0.17$  \\
S-III&  0.55  &$ 1.11  \pm 0.56 $&$ 0.63 \pm 0.20 $&$ -0.20  \pm 0.20 $&$ 0.20  \pm 0.12$\\ 
S-IV&  0.84  &$ 1.69  \pm 0.46 $&$ 0.07 \pm 0.20 $&$ -0.01  \pm 0.14 $&$ 0.19  \pm 0.12$ \\
S-V & 1.15& 1.74 $ \pm $ 0.67&  0.48  $ \pm $0.25 &-0.25 $ \pm $ 0.26&  0.08$ \pm $  0.20\\ 
  \hline
  
 \end{tabular}
 \label{kurtosis}
 \end{table*}

 \begin{table*}
\caption[]{ The skewness value of the stellar mass distribution of the BGGs in observations and SAMs which are computed with
the normal distribution. }

 \begin{tabular}{llllll}
 \hline\hline 
  Subsample ID & mean z & observations &   G11&  DLB07 &  B06\\
  \hline
S-I& 0.22   & $ -0.64 \pm 0.27 $ & $ -0.37 \pm 0.07 $ & $ -0.18 \pm 0.07 $ & $ -0.64 \pm 0.04$\\ 
S-II&  0.25   & $ -0.35 \pm 0.32 $ & $ -0.31 \pm 0.12 $ & $ -0.11  \pm 0.12 $ & $ -0.58  \pm 0.09$\\  
S-III &  0.55   & $ -0.77 \pm 0.26 $ & $ -0.28  \pm 0.09 $ & $ -0.28 \pm 0.10 $ & $ -0.58 \pm 0.06$\\ 
S-IV &  0.84   & $ -1.07 \pm 0.23 $ & $ -0.31 \pm 0.10 $ & $ -0.40 \pm 0.07 $ & $ -0.55 \pm 0.06$\\ 
S-V &  1.15   & $ -1.24 \pm 0.34 $ & $ -0.44 \pm 0.13 $ & $ -0.15 \pm 0.13 $ & $ -0.53 \pm 0.10$\\ 
  \hline
 \end{tabular} \label{skewness}
 \end{table*}

   \begin{figure*}
     \includegraphics[width=7cm]{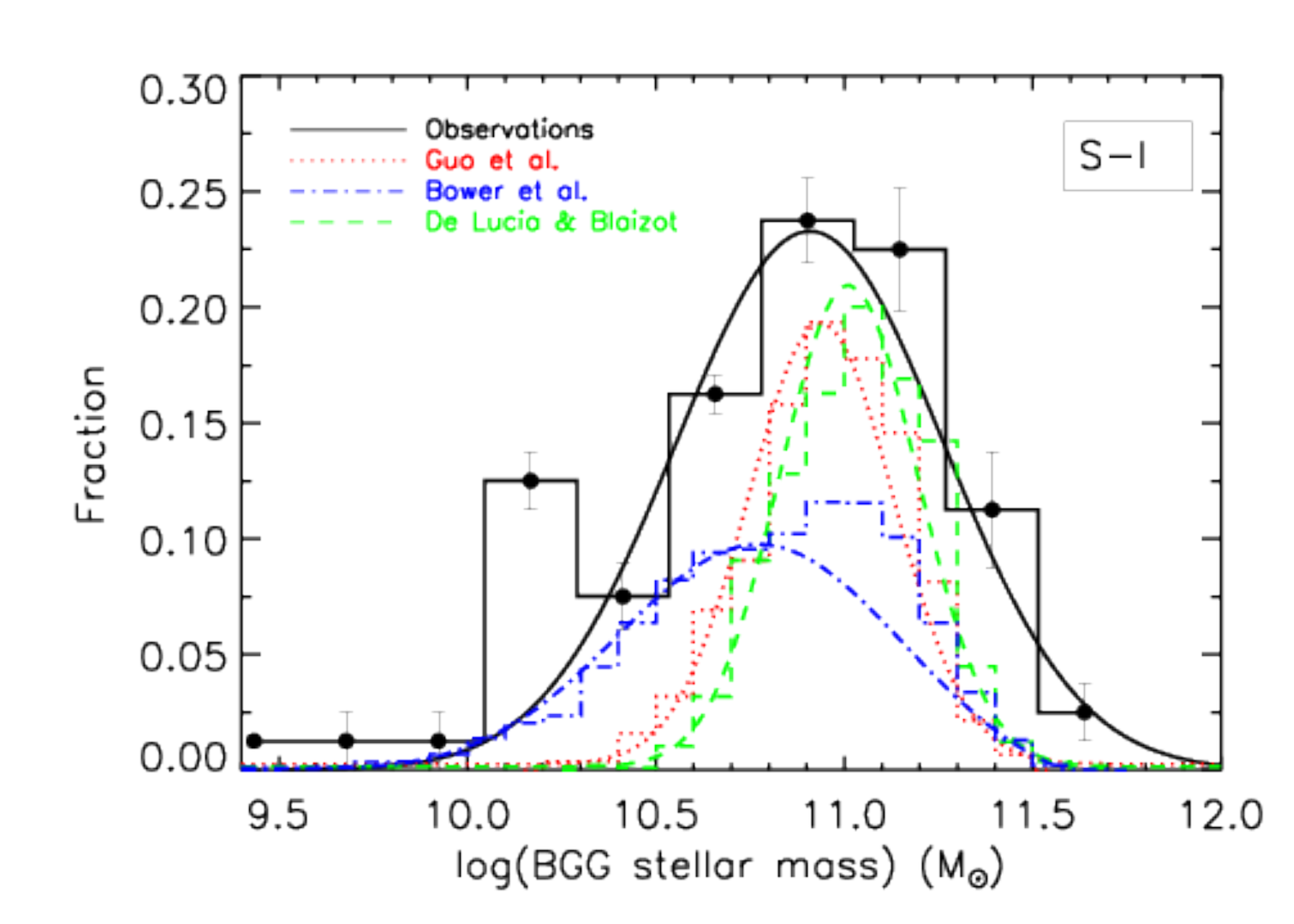}
     \includegraphics[width=7cm]{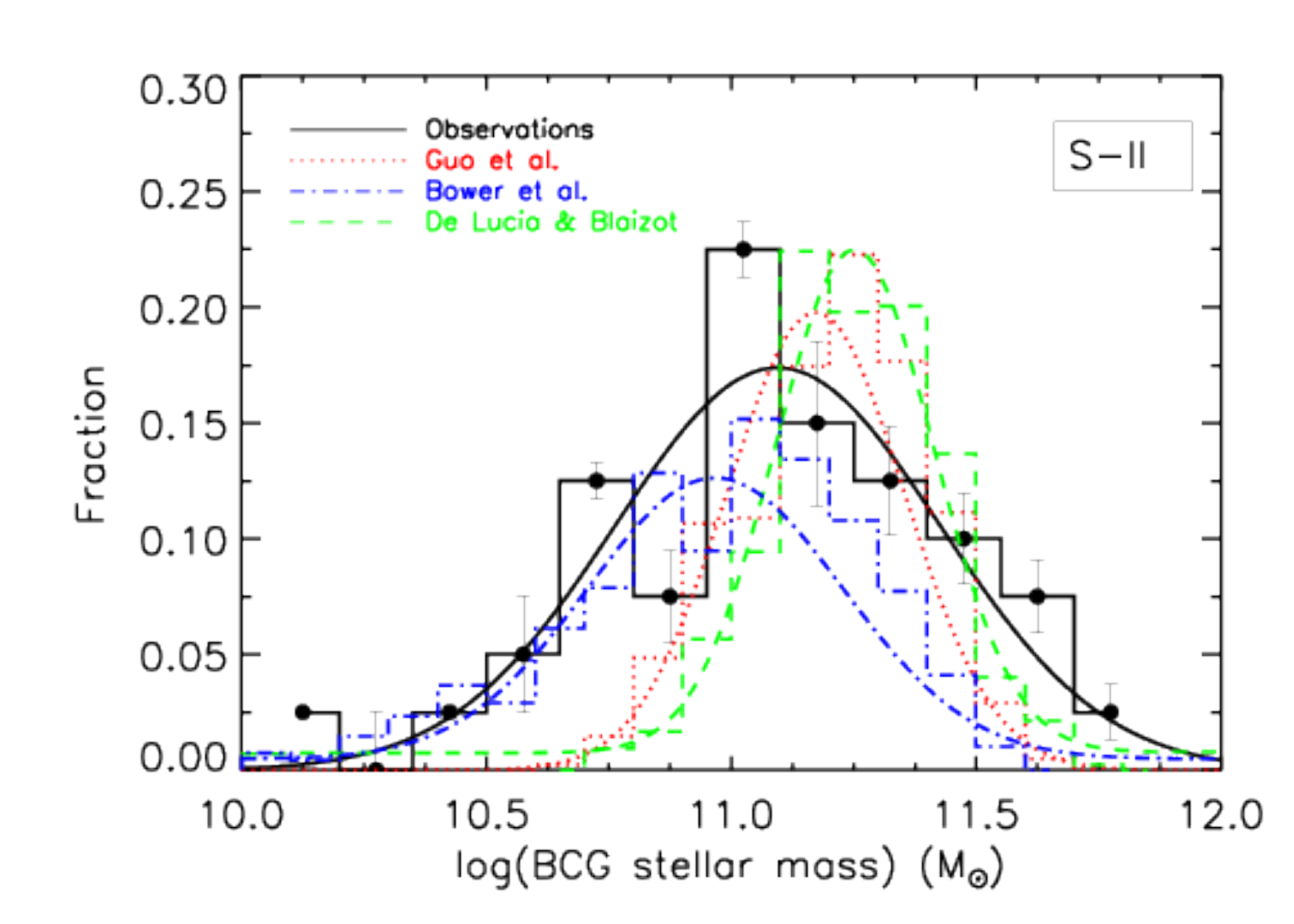}
     \includegraphics[width=7cm]{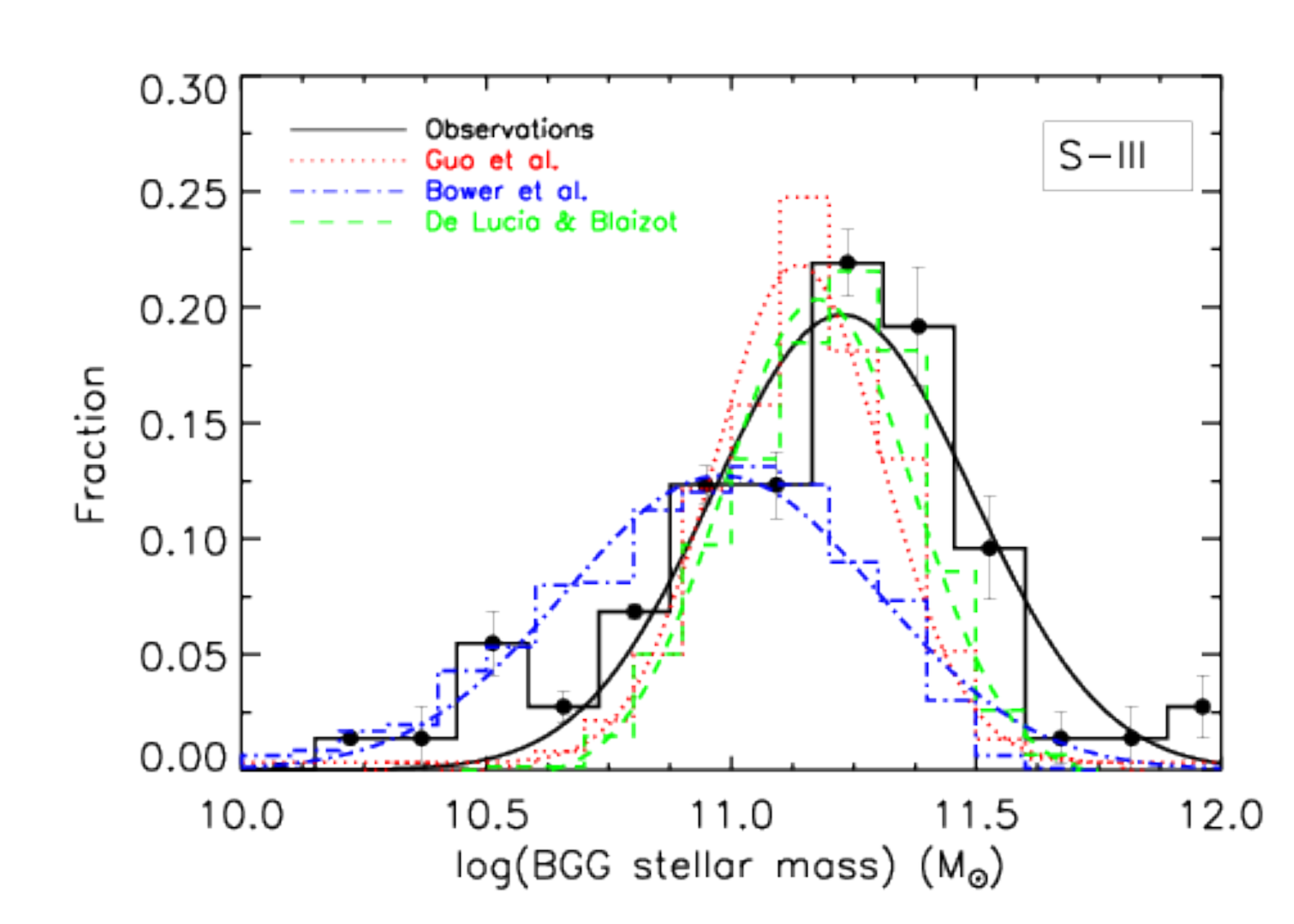}
     \includegraphics[width=7cm]{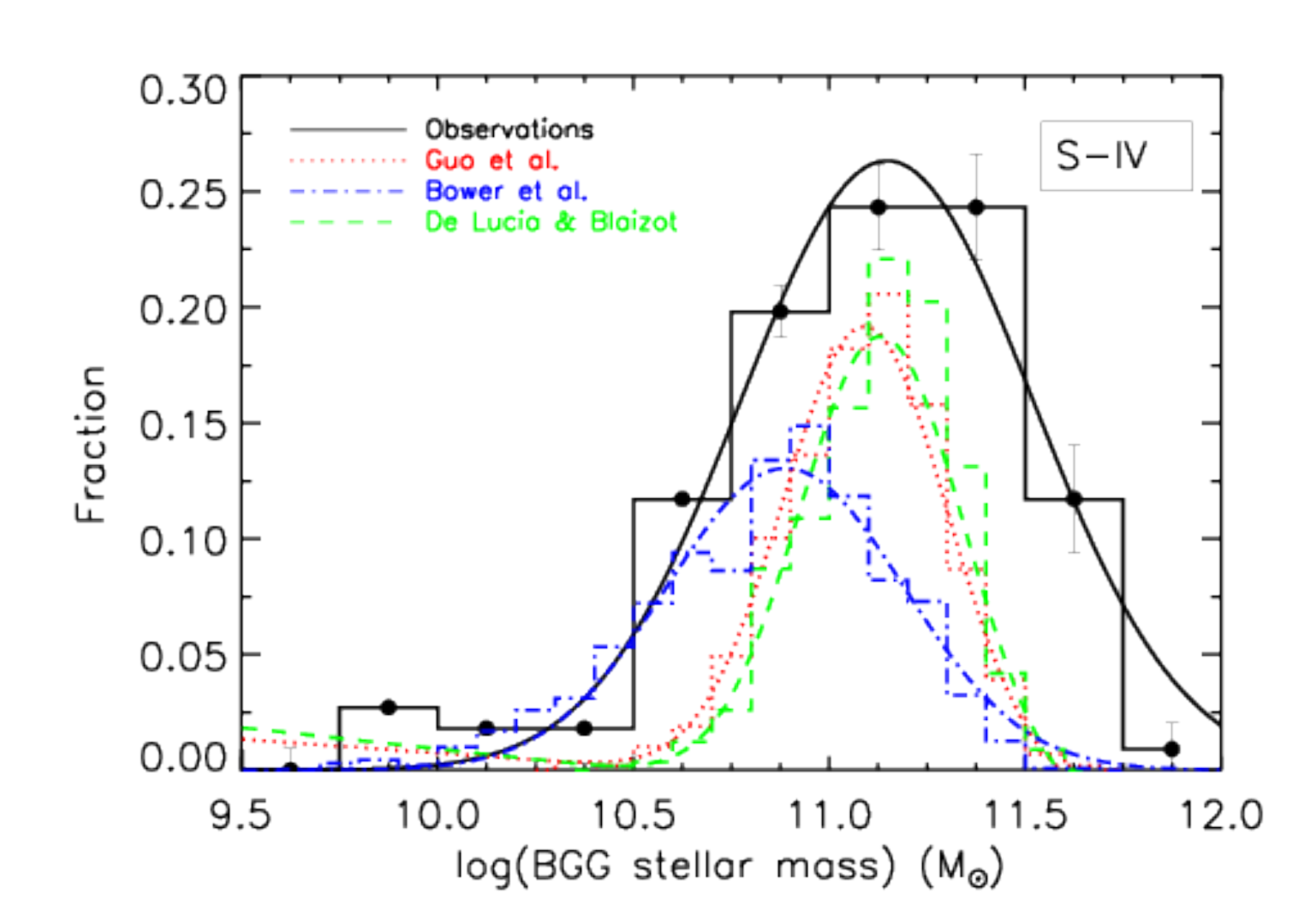}
       \includegraphics[width=7cm]{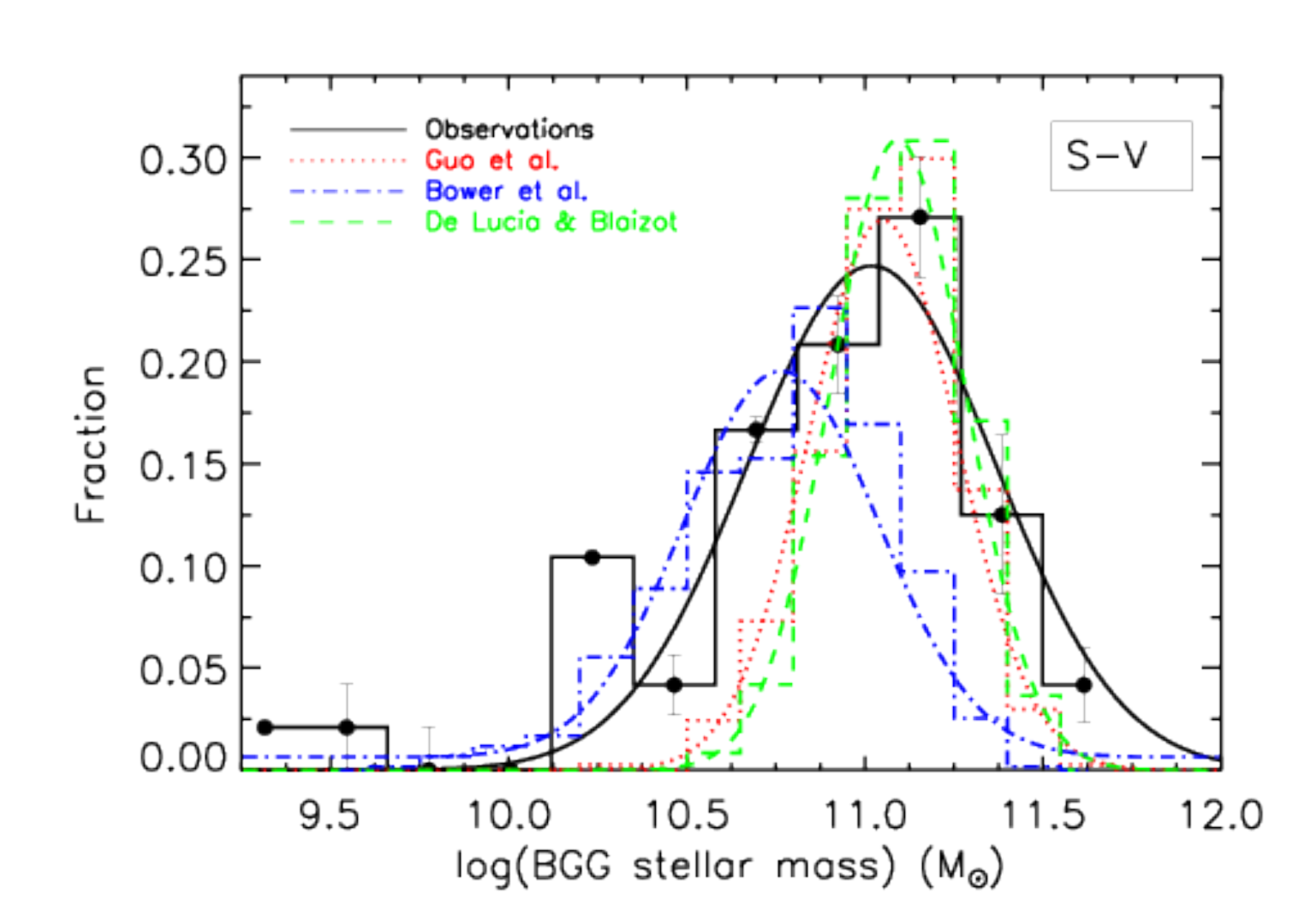}
    \includegraphics[width=7cm]{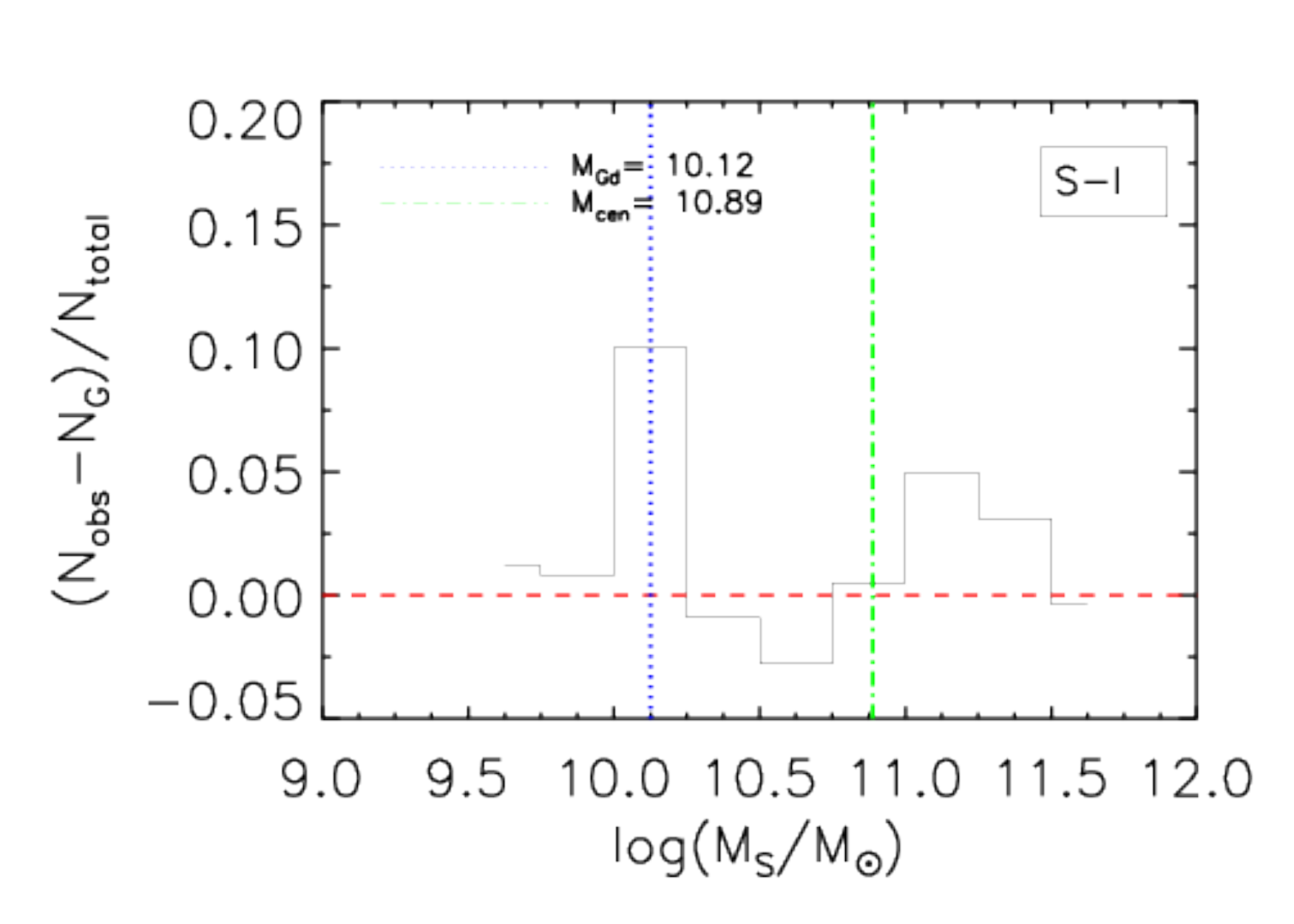}      
   \caption[]{The stellar mass distribution of the BGGs and Gaussian approximations are shown as solid black histogram and curve, respectively.  The predictions from the SAMs of G11, B06, and DLB07 are shown as the dotted red, dash-dotted blue, and dashed green histograms and curves, respectively. The lower--right panel shows the residual between the observed and the fitted fraction of the BGGs in similar stellar mass bins for S-I, the blue dotted and green dash-dotted lines correspond to the position of the maximum deviation ($ <log(M_{Gd})> $) of  the observed distribution from the Gaussian distribution and the position of the centre of the peak in the  Gaussian distribution ($log(M_{cen})$).}
     \label{sm_dis}
     \end{figure*}

\section {Evolution of the stellar mass} \label{results}
\subsection {Distribution of the stellar mass of the BGGs}\label{SMdis}
 
There are statistical studies of abundance matching, stacked lensing analysis, and clustering of galaxies that determine the mean occupation as a function of galaxy mass. These studies can not probe the distribution of galaxy properties, e.g.  stellar mass of galaxies for a given halo mass. The main advantage of our study is in the direct detection of galaxy groups which provides a unique opportunity to study the diversity of the BGG properties for a well defined sample in terms of the mass and the redshift of objects.

Four, out of five, subsamples (S-II to S-V) of galaxy groups cover a similar halo mass range. These  groups are rich galaxy groups and allow us to follow the evolutionary properties of their BGGs: the stellar mass distribution and the mass growth of the BGGs since $ z=1.3 $ to $ z=0.1 $. For each subsample, we construct the stellar mass distribution and quantify, in details, the shape
of  this distribution with respect to the normal distribution by fitting a single Gaussian distribution and measuring the skewness and the Kurtosis. We compare the best-fit Gaussian parameter, namely the center of the peak ($M_{cen}$), its height and the Gaussian width ($\sigma$), between the observations and semi-analytic models predictions. We also quantify the stellar mass bin, $M_{Gd}$ (Gd: Gaussian deviate), where the maximum deviation (e.g secondary peak) is occurred between the observed stellar mass distribution and the best-fit Gaussian distribution. The results are summarised in Table \ref{gg}.

In Fig. \ref{sm_dis}, we show the stellar mass distribution and the best-fit Gaussian function  to the observed data of BGGs (solid black histogram and curve) and that of the  SAM of G11 (dotted red histogram and curve), DLB07 (dashed green histogram and curve), and B06 (dash-dotted blue histogram and curve), respectively.

 The main findings are as follows:
 
I) The subsample S-V shows a secondary peak at $log(M_{Gd}/M_{\odot})\approx 10.2$. This peak is also seen in subsample S-III and S-II, however, the peak is slightly shifted towards higher mass bins.  These galaxies  with $log(M_{Gd}/M_{\odot})\approx 10.5$  are   a young population of BGGs. In a separate study we investigate the stellar age and star formation rate of this population.

II) The subsample S-I also shows a strong deviation from the single Gaussian fit as there appears to be a secondary peak at the low mass end. The peak is located in $log(M_{Gd}/M_{\odot})\approx 10.2$. We note that this subsample covers a lower halo mass compared to other subsamples.     
We  show the difference  between the fraction of BGGs in the observed distribution  ($N_{obs}$) and the best-fit Gaussian distribution  ($N_{G}$) in each stellar mass bin for S-I in the lower right panel in Fig. \ref{sm_dis}. The appearance of this secondary peak can also be suggestive of a newly forming BGGs. 

III) The observed distribution of the stellar mass covers a wide range. This appears to be only recovered by the B06 model which shows a similar Gaussian dispersion to the observational distribution of the stellar mass.  

IV) With a given similar stellar mass binsize, the peak value of  stellar mass distribution is over-predicted by  G11 and  DLB07 in all subsamples.

V) The location of the peak centroid is successfully reproduced by the G11 and the DLB07.

We find  that both G11 and DLB07 models perform similarly  in predicting the stellar mass distribution of the BGGs at all redshift and halo masses considered. However, the G11 prediction is  closer to observations than the DLB07 model. This model is an updated and modified version of the SAM which was presented in DLB07, and  both models share a number of prescription for the  physical processes which were used in their  implementations.  However, the comparison between our findings in observations and the G11 predictions indicates that this model still needs further development. 
\begin{figure}
\includegraphics[width=0.47\textwidth]{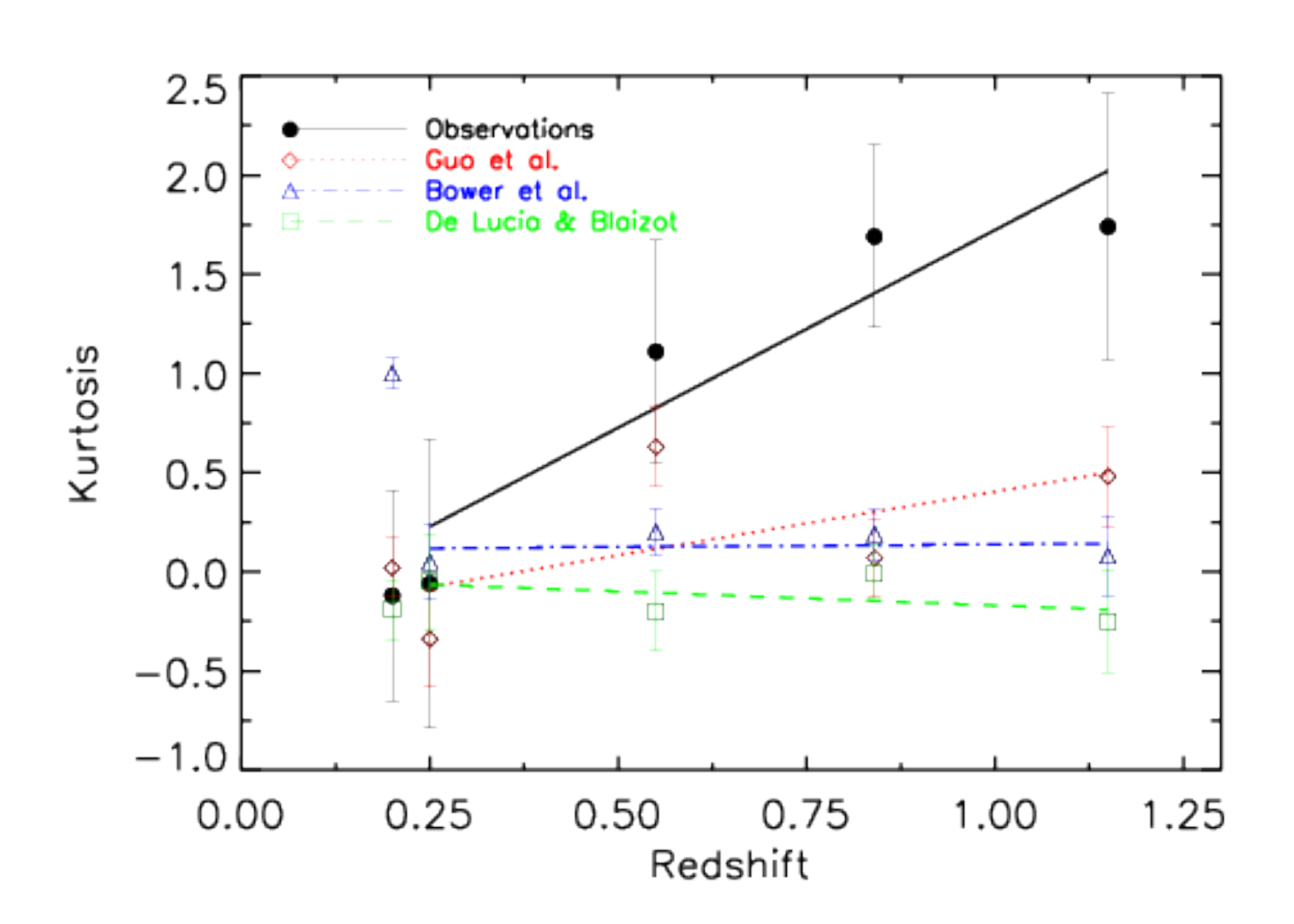}
 \includegraphics[width=0.47\textwidth]{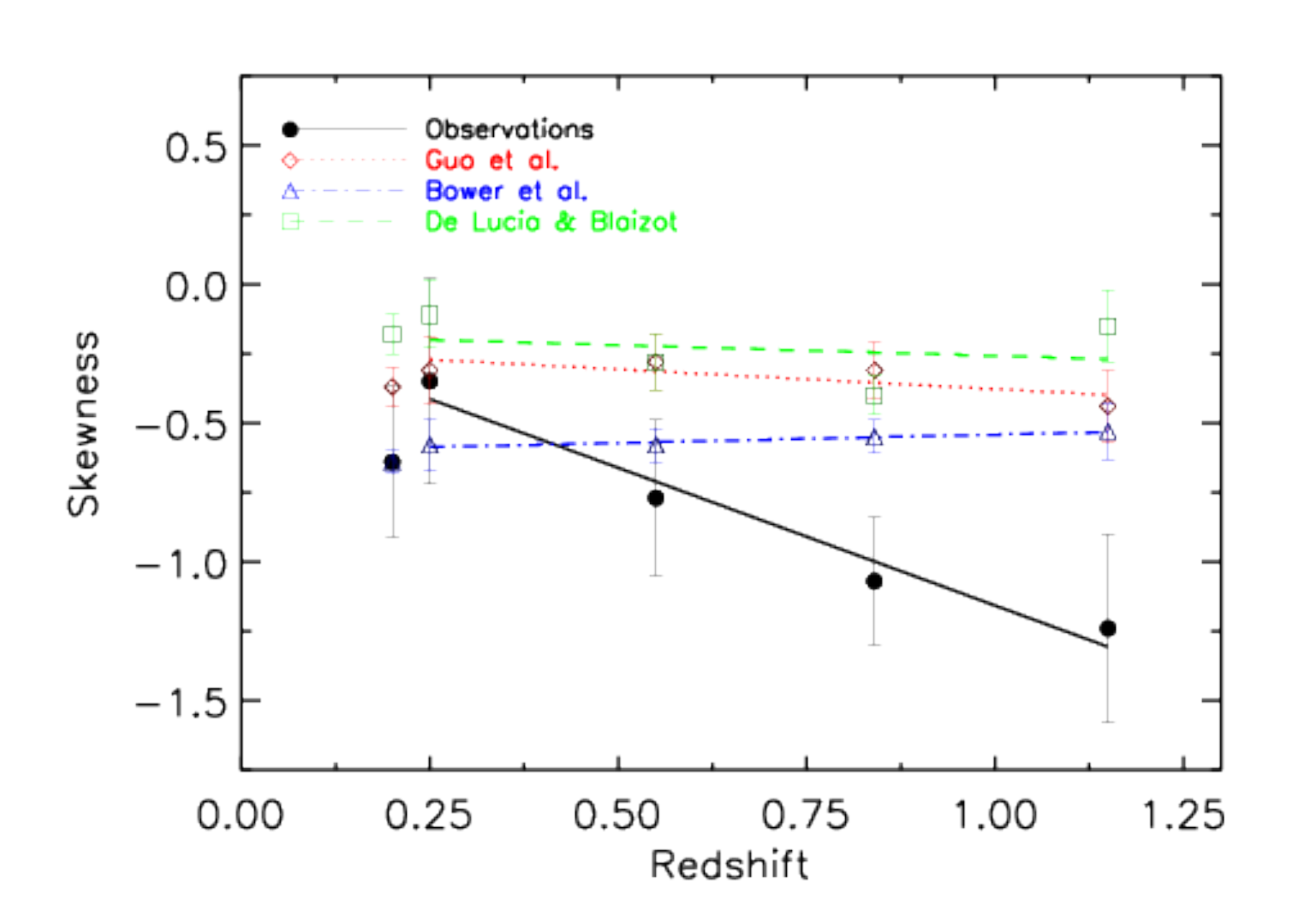}  
\includegraphics[width=0.47\textwidth]{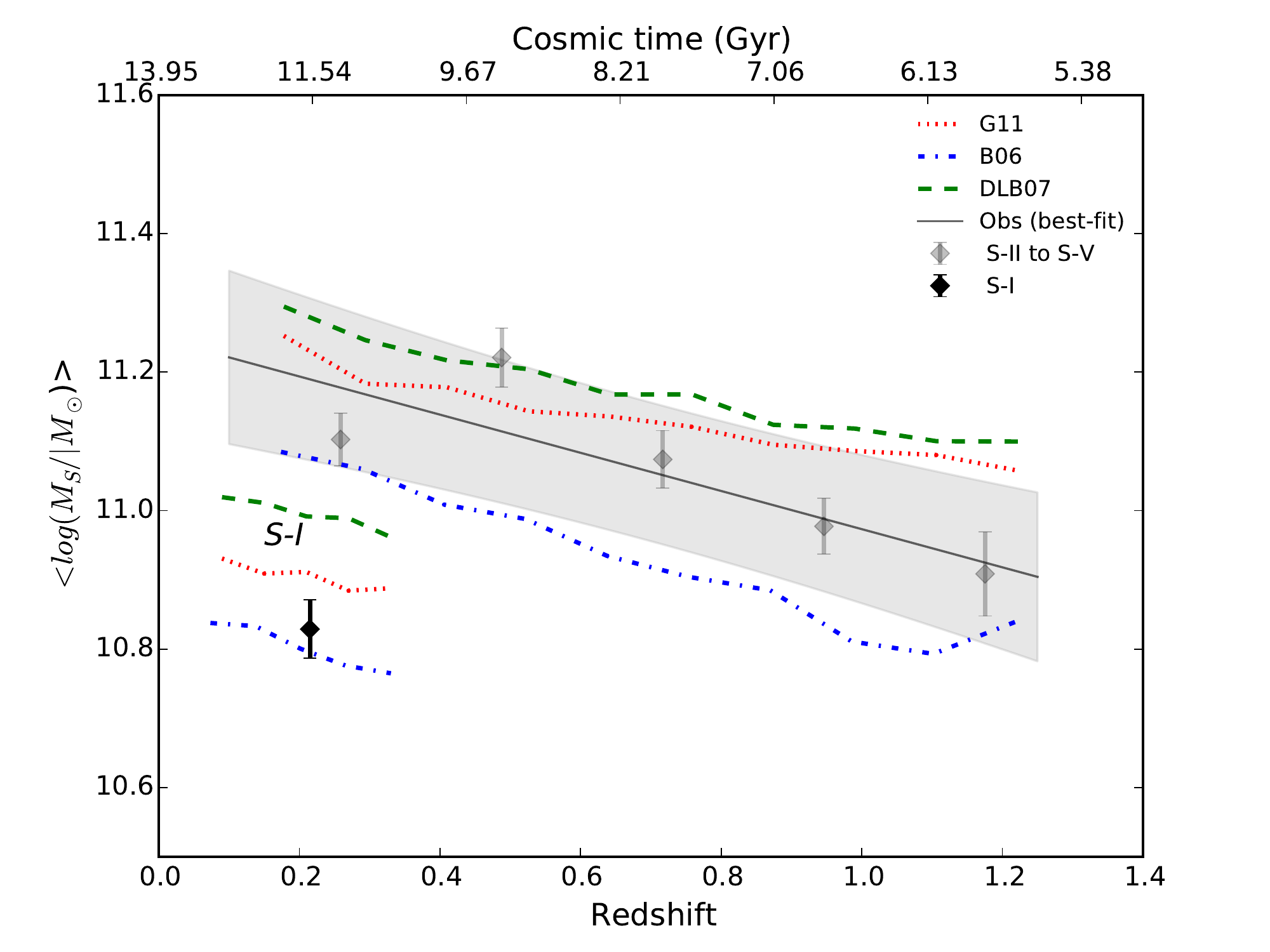}
\caption[]{The Kurtosis of the stellar mass distribution of the BGGs versus redshift (Upper panel) . The skewness of the stellar mass distribution of the BGGs versus redshift (Middle panel). The observed trend (solid black line) for the Kurtosis and skewness suggest that the shape of the stellar mass distribution evolves towards the normal distribution with decreasing redshift. The mean stellar mass of the BGGs versus redshift (Lower panel). The grey highlighted area represents the 68\% confidence interval of the best-fit to the data.The mean stellar mass of BGGs  grows  by a factor of $\sim2$ since $ z=1.3 $ to present day. }
\label{z-skewpeak}
\end{figure}

Moreover, the discrepancies between observations and the three SAMs in particular at low masses can possibly be linked to the star forming activities  and the mass assembly  history of the BGGs.  For example, SAMs generally overestimate the number of the low mass satellite galaxies, thus BGGs in models can possibly experience more minor mergers than the BGGs in observations.  Recently, \cite{Cousin15}, constructed four SAMs, which three of them are classical SAMs and the $ 4^{th} $ SAM is a model which is constructed based on a $ \Lambda CDM $ model with a correct number of the low-mass halos and adopting   two-phases gaseous disk: the first with the star-forming gas and the second with no-star-forming gas. They showed  that even when a strong SN-feedback and photoionisation are applied on the implementation of  three classical SAMs, these models  again form too many stars and over-predict the low mass end of the stellar mass function. However, the $ 4^{th} $ SAM is matched to the observations very well. They argue that at any given redshift, only a fraction of the total gas in a galaxy is used for the star formation activity.

\subsection{Evolution of the shape of the stellar mass distribution} \label{sp}
In section \S \ref{SMdis}, we find a secondary peak in the stellar mass distribution at  low mass tail, a feature that are not predicted by SAMs. The observed Gaussian parameters also seem to vary with redshift.  In order to quantify the redshift evolution of the shape of the stellar mass distribution, we measure the Kurtosis and skewness values of the stellar mass distribution in each subsample as presented in Tab. \ref{kurtosis} and Tab. \ref{skewness}.

The upper and middle panels of Fig. \ref{z-skewpeak} shows  the Kurtosis and skewness values of the stellar mass distribution of the BGGs versus the mean redshift of each subsample and linear fit. The observed trend is shown with  the solid black line and the predicted trends have been shown with dotted red line (G11), dashed green line (DLB07), and dot-dashed blue line (B06), respectively. The non-connected symbols present results for S-I.  The skewness of the stellar mass distribution tends to decrease with redshift. The negative values of this parameter indicates that the stellar mass distribution tends to skew more towards low masses. The Kurtosis value considerably increases with redshift indicating that the stellar mass distribution becomes progressively peaked.  The predicted Kurtosis and skewness values by SAMs shows no or a little evolution. The B06 model has a best consistency with the data below $ z\sim 0.5 $, but above this redshift all models fail to predict the observed stellar mass distribution of the BGGs.

Fig. \ref{z-skewpeak} clearly demonstrates that the shape of the stellar mass distribution in observations significantly evolves towards a normal distribution with decreasing redshift. The shape of the stellar mass distribution in model predictions shows no evolution  and this  distribution approximately follows a normal distribution at all redshifts.

\subsection{Evolution of the mean stellar mass of  BGGs }\label{msm}   
 The lower panel of Fig. \ref{z-skewpeak} shows the evolution of the mean stellar mass of the BGGs  in observations (filled  dimonds) over the redshift range of $ 0.1<z<1.3 $.  To quantify the growth of the mean mass of BGGs we re-sample BGGs within massive halos  ($M_{200}=10^{13.5} $ to $ 10^{14} M_{\odot}$)  into 5 redshift bins. This helps to model the mean mass evolution more accurately. The solid gray line  and the highlighted area indicate the best-fit linear relation ($<log(M_{S})>=(-0.27\pm 0.1)z+(11.24\pm0.08)$) to the  observed data and its  $ (68\%) $ confidence intervals, respectively. Using the best-fit function  we find that the mean stellar mass of BGGs grows gradually by a factor of $ \sim2.07 $ since $z=1.3 $ to $z=0.1$.

On the lowest panel of Fig.  \ref{z-skewpeak}, we also show the evolution of the mean stellar mass of BGGs in the SAM  predictions (G11, DLB07, and B06). The modeled trends are approximated by  the best-fit linear relations as follows (since models predict roughly linear trends, thus  we ignore to plot these relations):\\
 
B06 :

 $<log(M_{S})>= (-0.28\pm 0.02)z+(11.13\pm0.02) $\\

DLB07 :

  $<log(M_{S})>= (-0.18\pm 0.01)z +  (11.3\pm0.01)$\\

G11:

$<log(M_{S})>= (-0.17\pm 0.01)z +  (11.25\pm0.01)$\\

The mean stellar mass of BGGs in the SAM evolves by a factor of 2.09 (B06), 1.63 (DLB07) and 1.55 (G11) since $z=1.3$ to $ z=0.1$, respectively.  Evolution of the mean stellar mass of BGGs in all models is consistent with that of observations within the errors. Among models, the rate of stellar mass evolution in B06  is the closest to observations. 

We note that there is some tension between model predictions and observations at $z<0.4$, as the modeled slope of the stellar  mass growth is steeper than that in the observations. In addition,  the observed growth of the BGG mass mainly occurs at $0.4<z<1.3$. Within observational errors, our findings are  consistent with that of early studies e.g. \cite{Lidman12}. However, \cite{Lidman12} have studied the BCG assemblies  within massive halos and found that these object can grow in mass by a factor of $ 1.8\pm 0.3 $ at $ 0.1<z<0.9 $. In addition, The lack of the mass growth at low redshifts ($ z<0.5 $) has also been argued by \cite{Oliva14} and \cite{Lin13}. 

In lower panel in Fig. \ref{z-skewpeak},  we separately show the mean mass of the BGGs in observations (filled black diamond) and models for S-I. DLB07 over-predicts the mean stellar mass of the BGGs for S-I, while B06 and G11 agree with observations. 
    
 \begin{figure*}
    \includegraphics[width=0.48\textwidth]{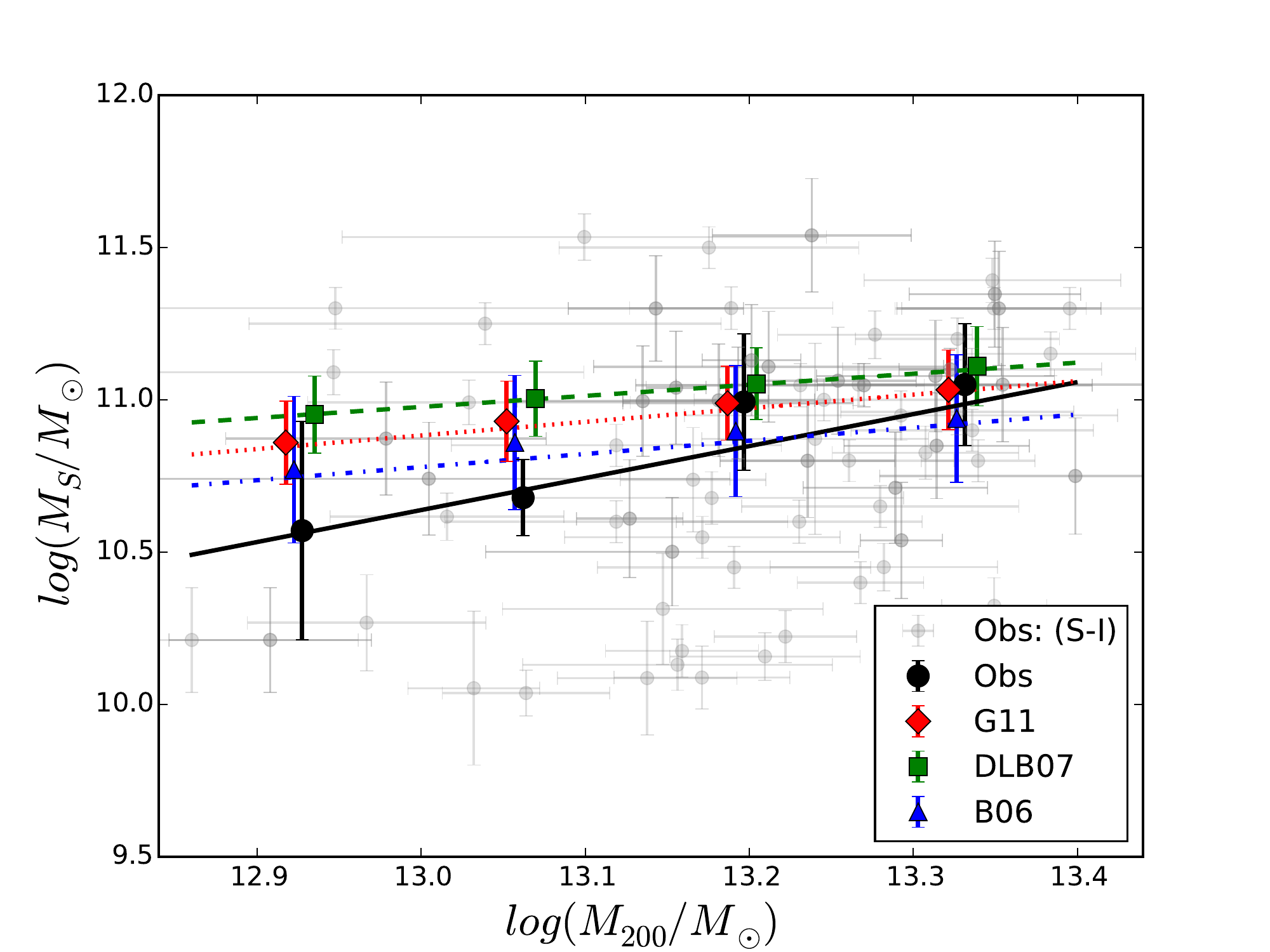}
    \includegraphics[width=0.48\textwidth]{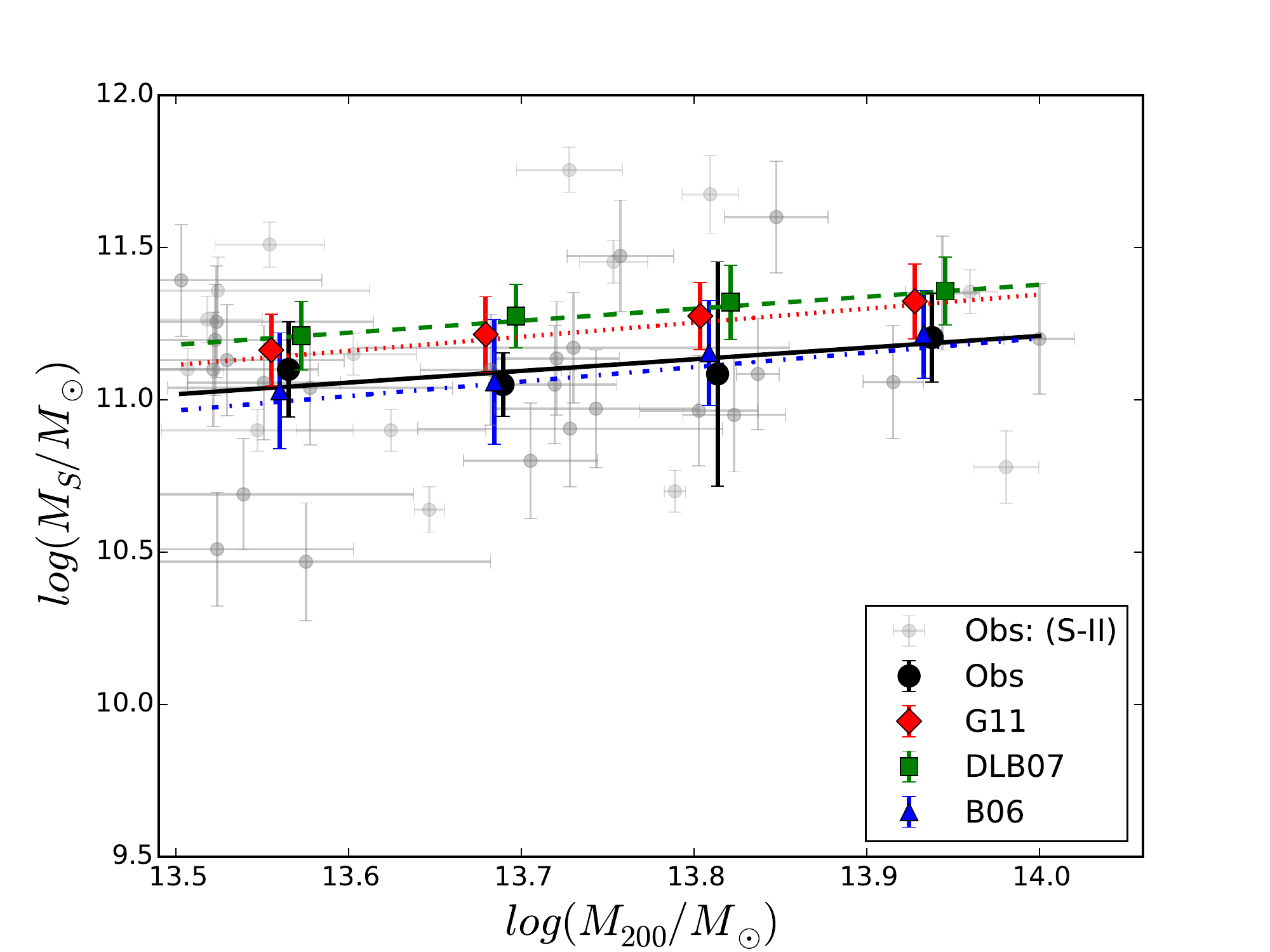}
    \includegraphics[width=0.48\textwidth]{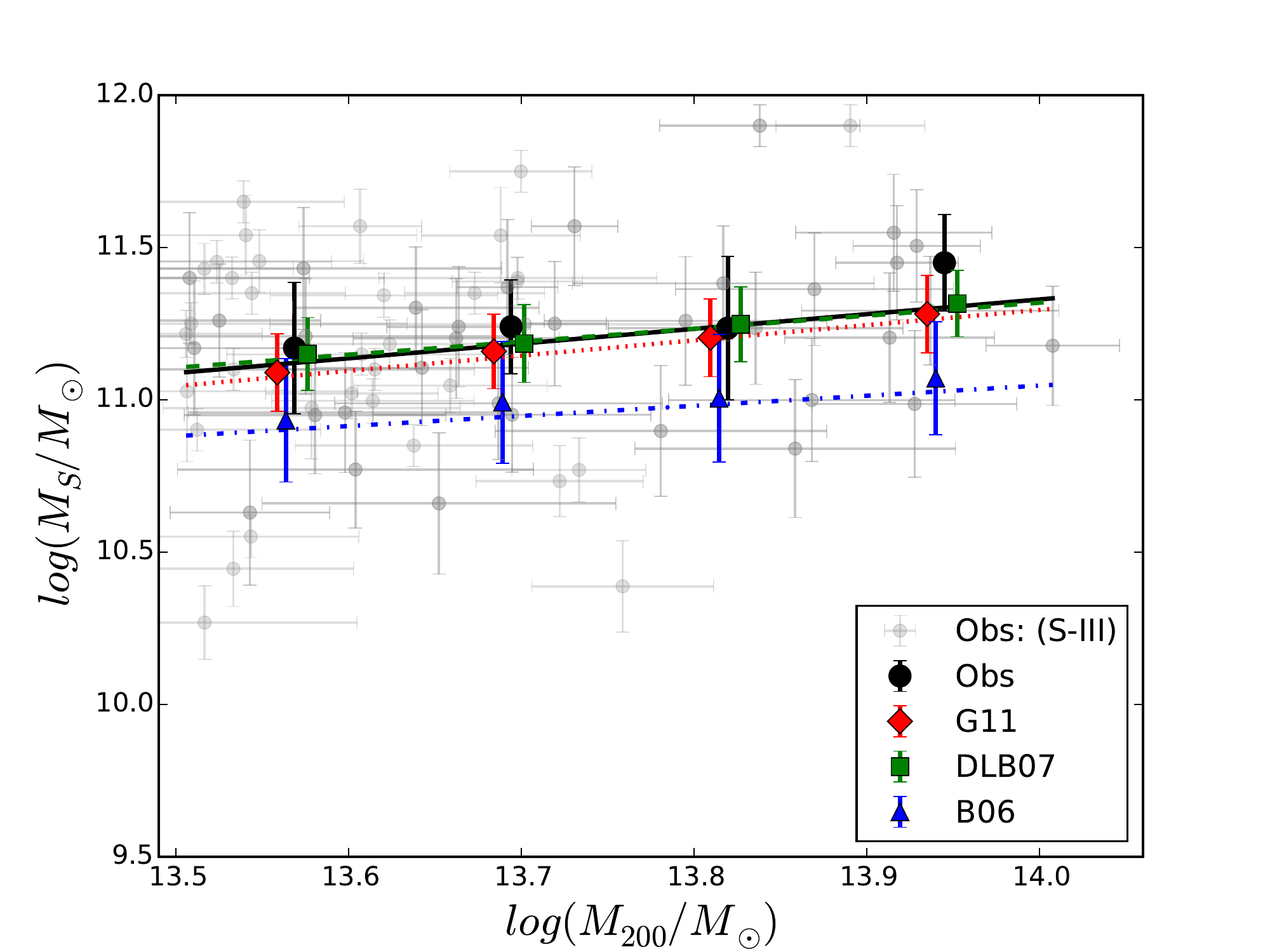}
    \includegraphics[width=0.48\textwidth]{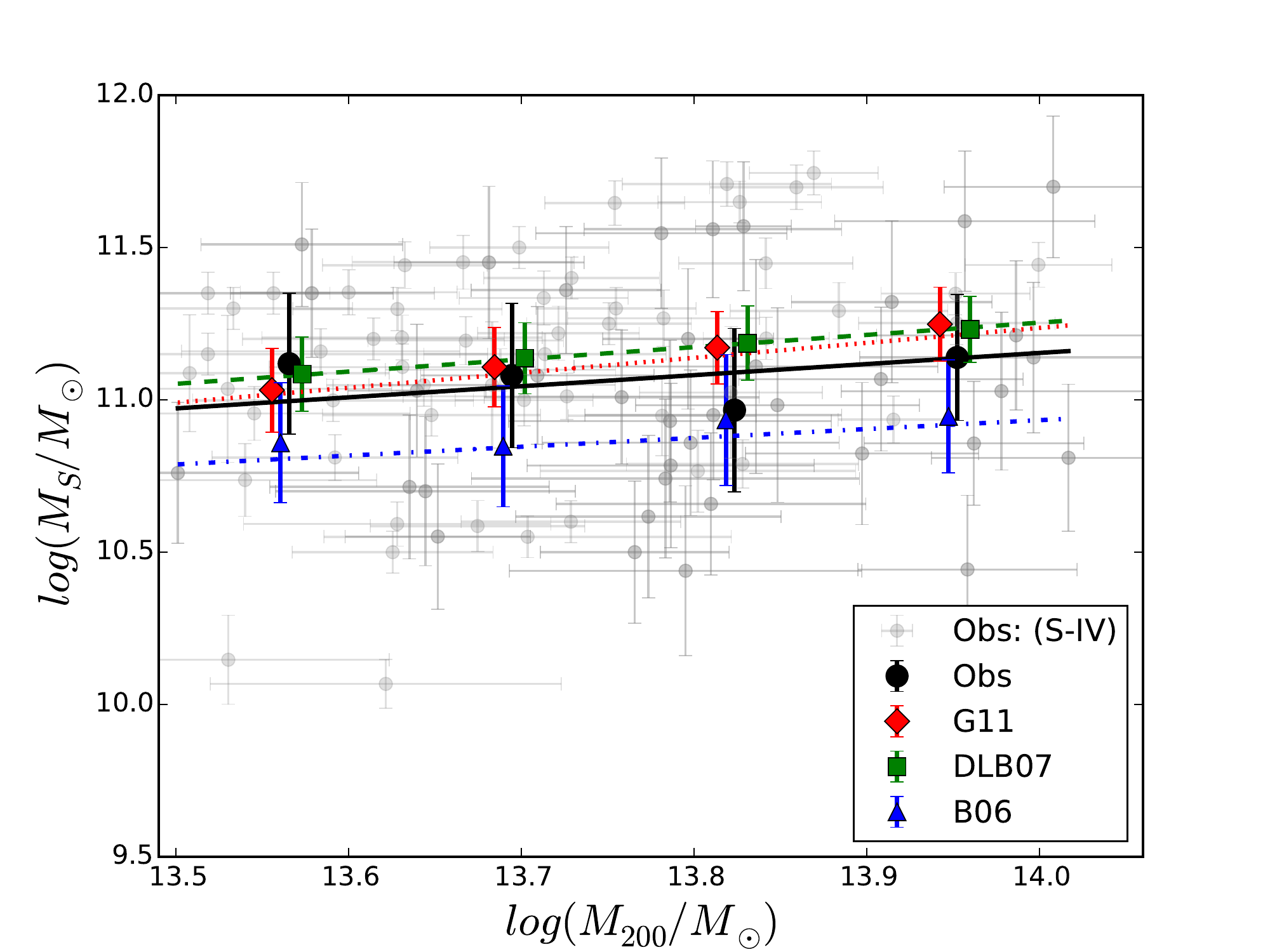}
    \includegraphics[width=0.48\textwidth]{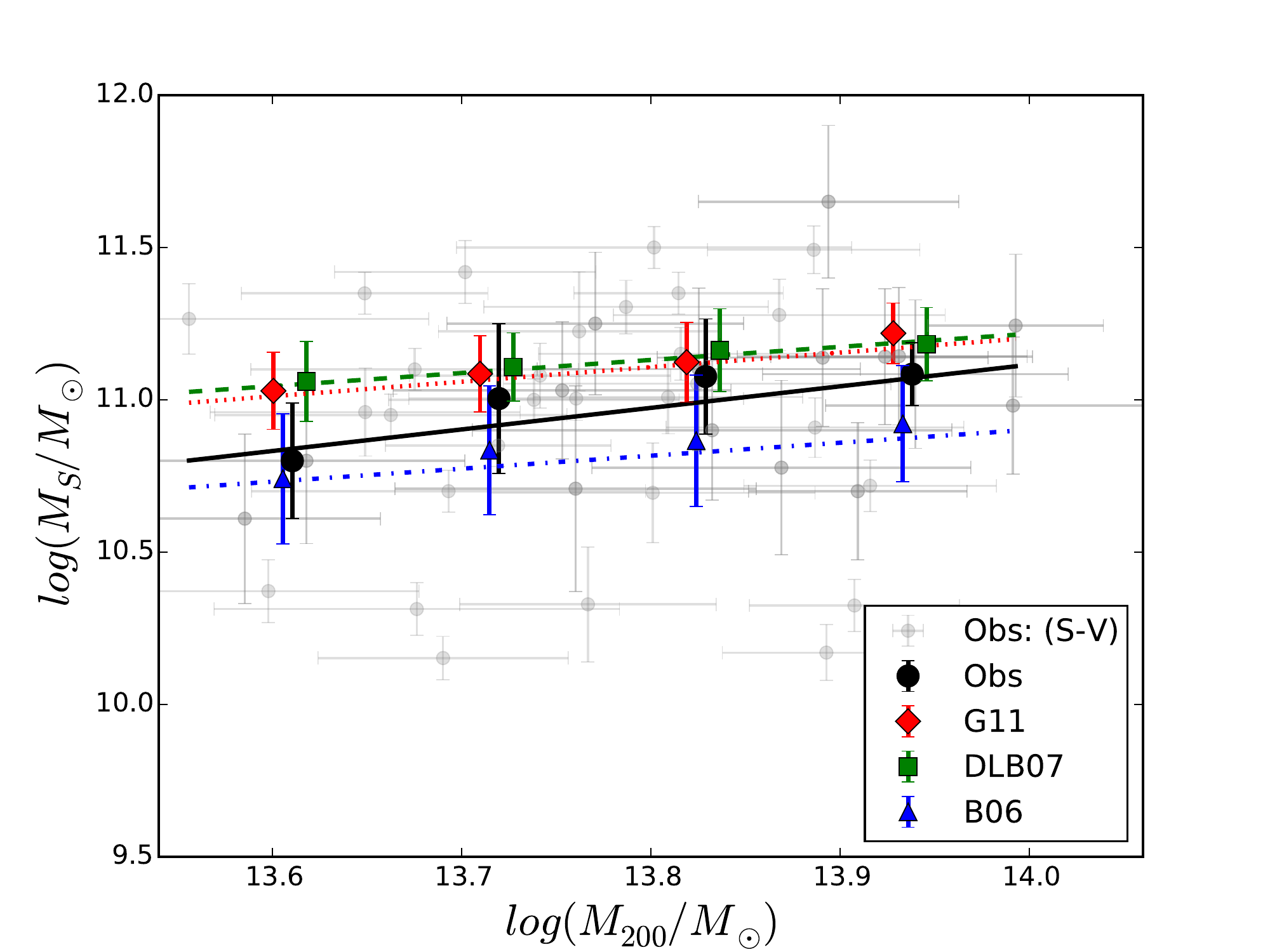}
  \caption[]{The stellar mass of  BGGs against the group mass ($M_{200}$)  in observations (gray points) for S-I (top-left panel), S-II (top-right panel), S-III (middle-left panel) and S-IV (middle-right panel), and S-V (bottom panel), respectively. The dark circles and line, red diamonds and dotted line, green squares and dashed line, and blue triangles and dash-dotted line show the median trends and the best-fit relation in observations and the SAMs of G11, DLB07, and B06, respectively. In agreement with model predictions  the BGG mass is positively correlates with the  host group mass.} 
  \label{m200Sm}
   \end{figure*}       
\subsection{The stellar mass and halo mass relation} 

 We study the correlation between the stellar mass of the BGGs and the total mass  ($ M_{200} $) of their host groups in observations  and SAMs. We show the results in  Fig. \ref{m200Sm}. The median value of the stellar mass of BGGs with the associated error (median absolute deviation) in each halo mass bin is also presented in this figure. Furthermore, we apply a robust linear regression and estimate the best fit relation ($ log (M_{S})= \beta \times log(M_{200})+ \alpha $ ) over the whole data in both observations and SAMs, presented in Tab.  \ref{m200t}. 
   
Our major findings  are as follows: in accordance with model predictions, the trends  show that the stellar mass of BGGs  positively correlates with their host group mass at all redshifts, indicating that the massive halos host massive  BGGs. The observed median trend of the $ M_{S}-M_{200} $ relation  is consistent  with model predictions within errors in all subsamples. We point out that  the best-fit relation  in observations for S-I is steeper than that of the model predictions and the observed trend  for S-II. This indicates that the stellar mass of BGGs within low mass groups correlates more strongly  with the halo mass compared to that of  BGGs within massive groups  at similar redshift range ($0.04<z<0.4$).  
 
We find that the observed best-fit relations for high-z subsamples (S-II to S-V) are almost similar and consistent with model predictions within uncertainties. 
\begin{table}
\caption[]{ \btxt{The best fit relation ($ log (M_{S})= \beta \times log(M_{200})+ \alpha $ ) on the stellar mass and halo mass plane in observations and SAMs.}}
 \begin{tabular}{lll} 
 \hline\hline  
  Subsamples& $ \beta $ & $  \alpha $  \\
  \hline
\textbf{S-I:}\\
 Obs & 1.05 $\pm$ 0.3 & -3.01 $\pm$ 3.9\\
G11 & 0.45 $\pm$ 0.02 & 5.08 $\pm$ 0.23\\
B06 & 0.43 $\pm$ 0.03 & 5.18 $\pm$ 0.38\\
DLB07 & 0.36 $\pm$ 0.02 & 6.26 $\pm$ 0.21\\
\hline
\textbf{S-II:}\\
Obs & 0.38 $\pm$ 0.24 & 5.85 $\pm$ 3.26\\
G11 & 0.46 $\pm$ 0.04 & 4.88 $\pm$ 0.56\\
B06 & 0.47 $\pm$ 0.07 & 4.56 $\pm$ 0.92\\
DLB07 & 0.39 $\pm$ 0.04 & 5.87 $\pm$ 0.52\\
\hline
\textbf{S-III:}\\
Obs & 0.48 $\pm$ 0.21 & 4.56 $\pm$ 2.89\\
G11 & 0.5 $\pm$ 0.02 & 4.3 $\pm$ 0.27\\
B06 & 0.33 $\pm$ 0.03 & 6.4 $\pm$ 0.43\\
DLB07 & 0.43 $\pm$ 0.02 & 5.35 $\pm$ 0.26\\
\hline
\textbf{S-IV:}\\
Obs & 0.36 $\pm$ 0.23 & 6.06 $\pm$ 3.12\\
G11 & 0.49 $\pm$ 0.03 & 4.36 $\pm$ 0.42\\
B06 & 0.29 $\pm$ 0.05 & 6.87 $\pm$ 0.65\\
DLB07 & 0.4 $\pm$ 0.03 & 5.63 $\pm$ 0.37\\
\hline
\textbf{S-V:}\\
 Obs & 0.71 $\pm$ 0.37 & 1.2 $\pm$ 5.11\\
G11 & 0.48 $\pm$ 0.05 & 4.5 $\pm$ 0.72\\
B06 & 0.43 $\pm$ 0.08 & 4.94 $\pm$ 1.15\\
DLB07 & 0.43 $\pm$ 0.05 & 5.2 $\pm$ 0.66\\
  \hline
  
 \end{tabular}
 \label{m200t}
 \end{table}

Previous studies have also shown that the stellar mass of the BGGs positively correlates with the hosting halo masses. For
example, \cite{stott10}  have shown that the stellar mass of the BGGs within their sample of 20 massive clusters at $0.8< z< 1.5$ is correlated with cluster mass. Recently, \cite{Oliva14} also examined the $ M_{S}-M_{200} $ relationship for both centrally and non-centrally located BGGs selected from the Galaxy And Mass Assembly (GAMA) survey  at $ z=0.09-0.27$ and found that for both subsamples the $M_{S}-M_{200} $ relation follows a power low $ (\sim0.32\pm0.2) $.

  \begin{figure*}
 \includegraphics[width=0.48\textwidth]{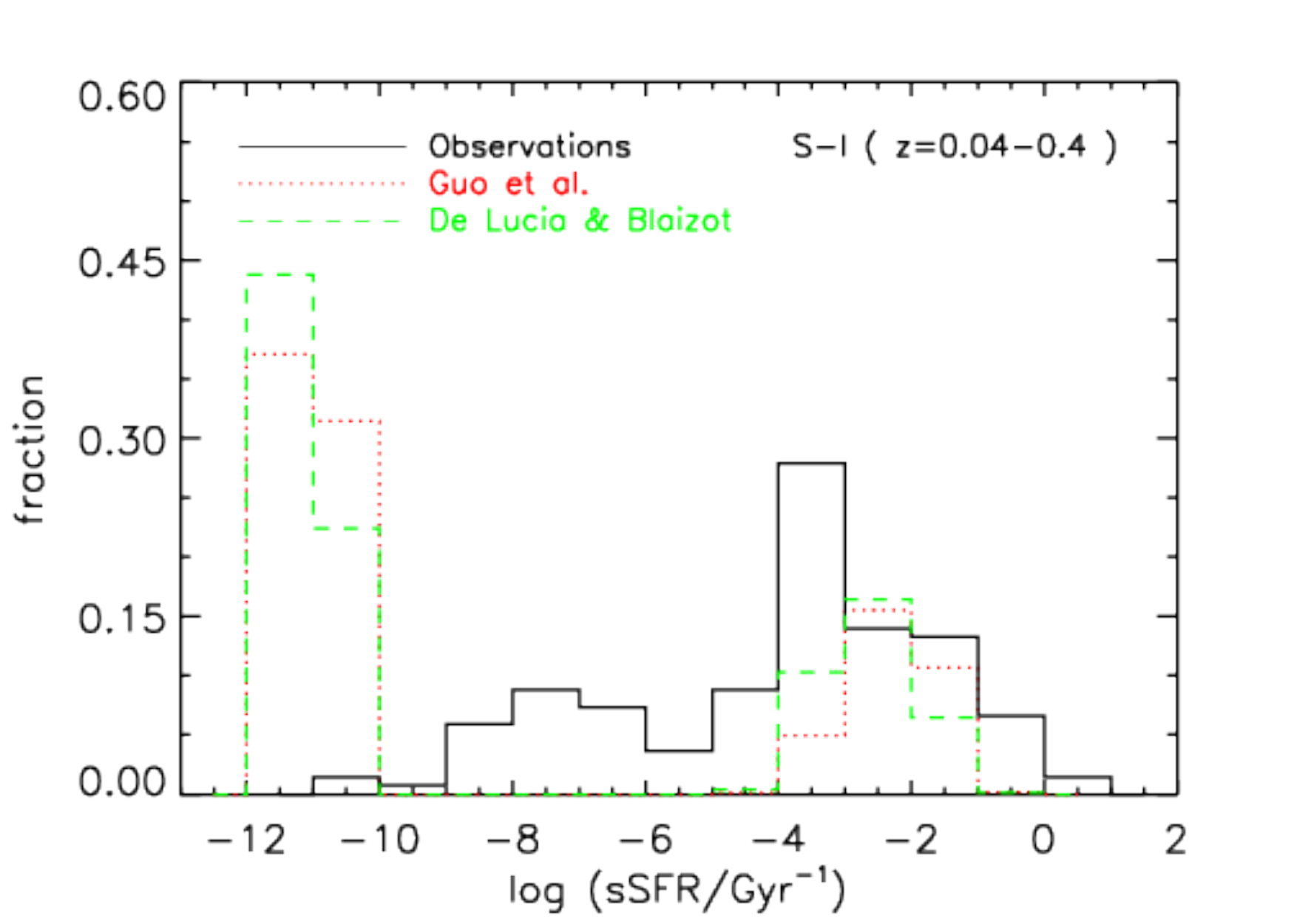}
  \includegraphics[width=0.48\textwidth]{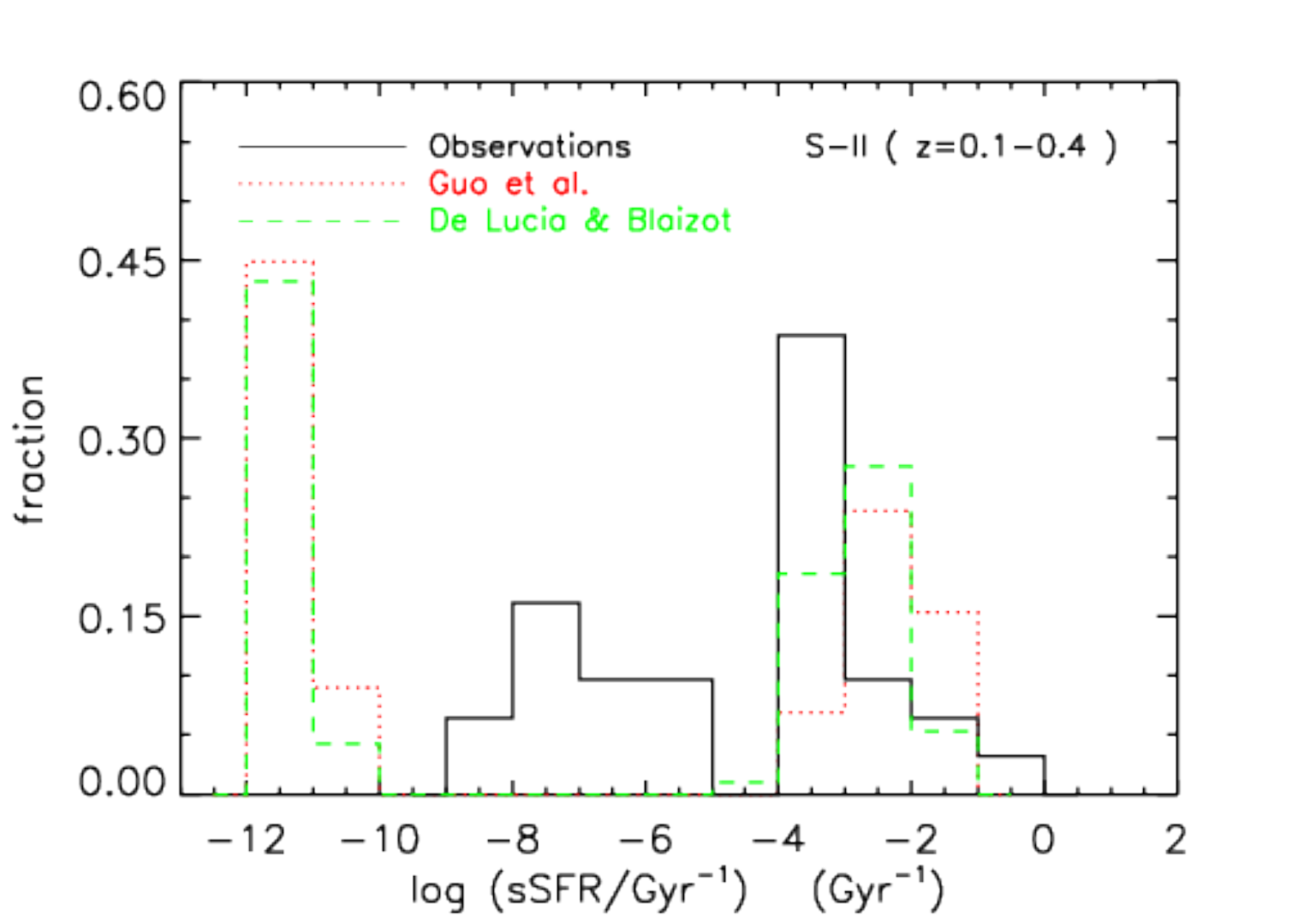}
  \includegraphics[width=0.48\textwidth]{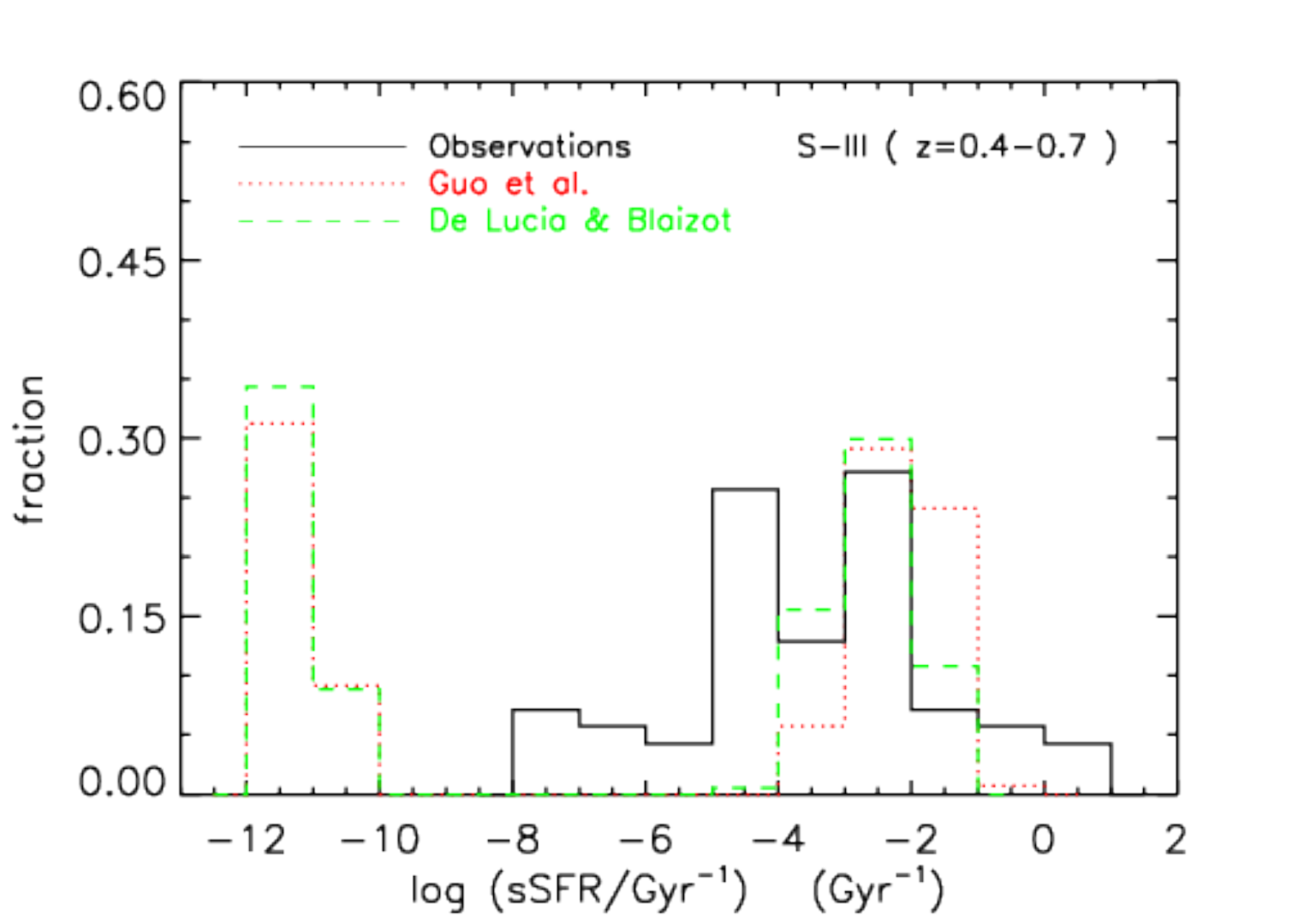}  
 \includegraphics[width=0.48\textwidth]{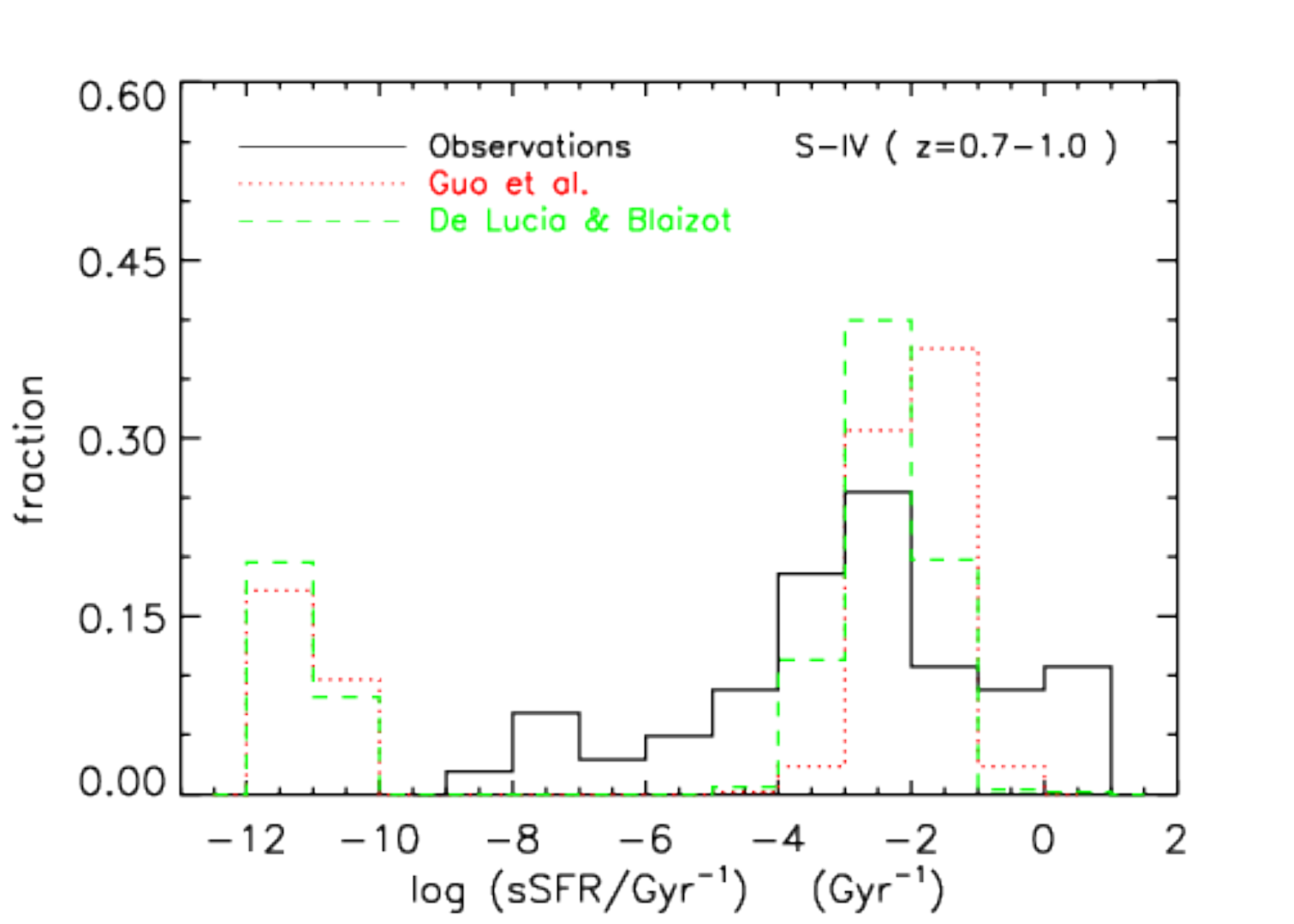}  
 \includegraphics[width=0.48\textwidth]{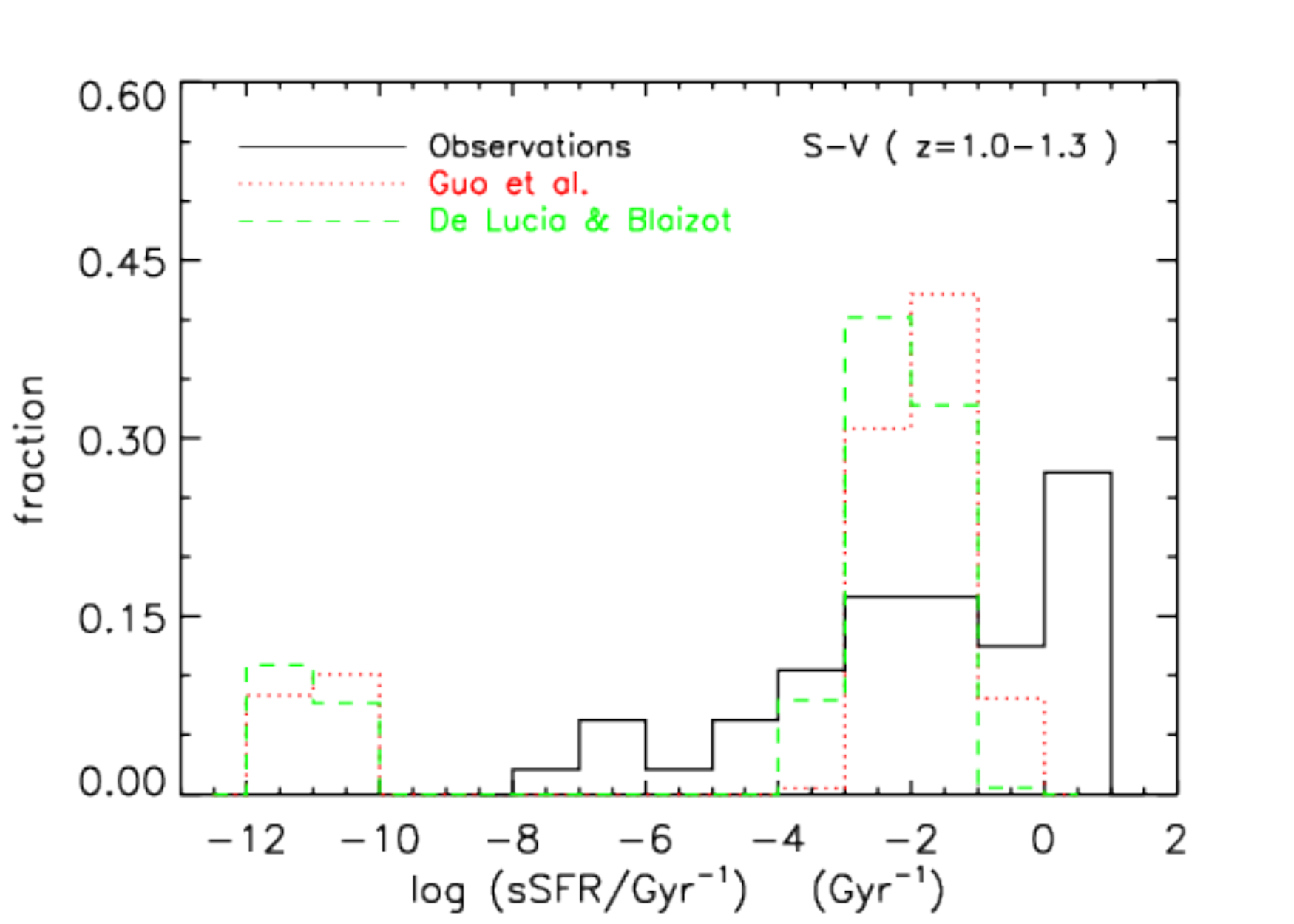}  
\includegraphics[width=0.48\textwidth]{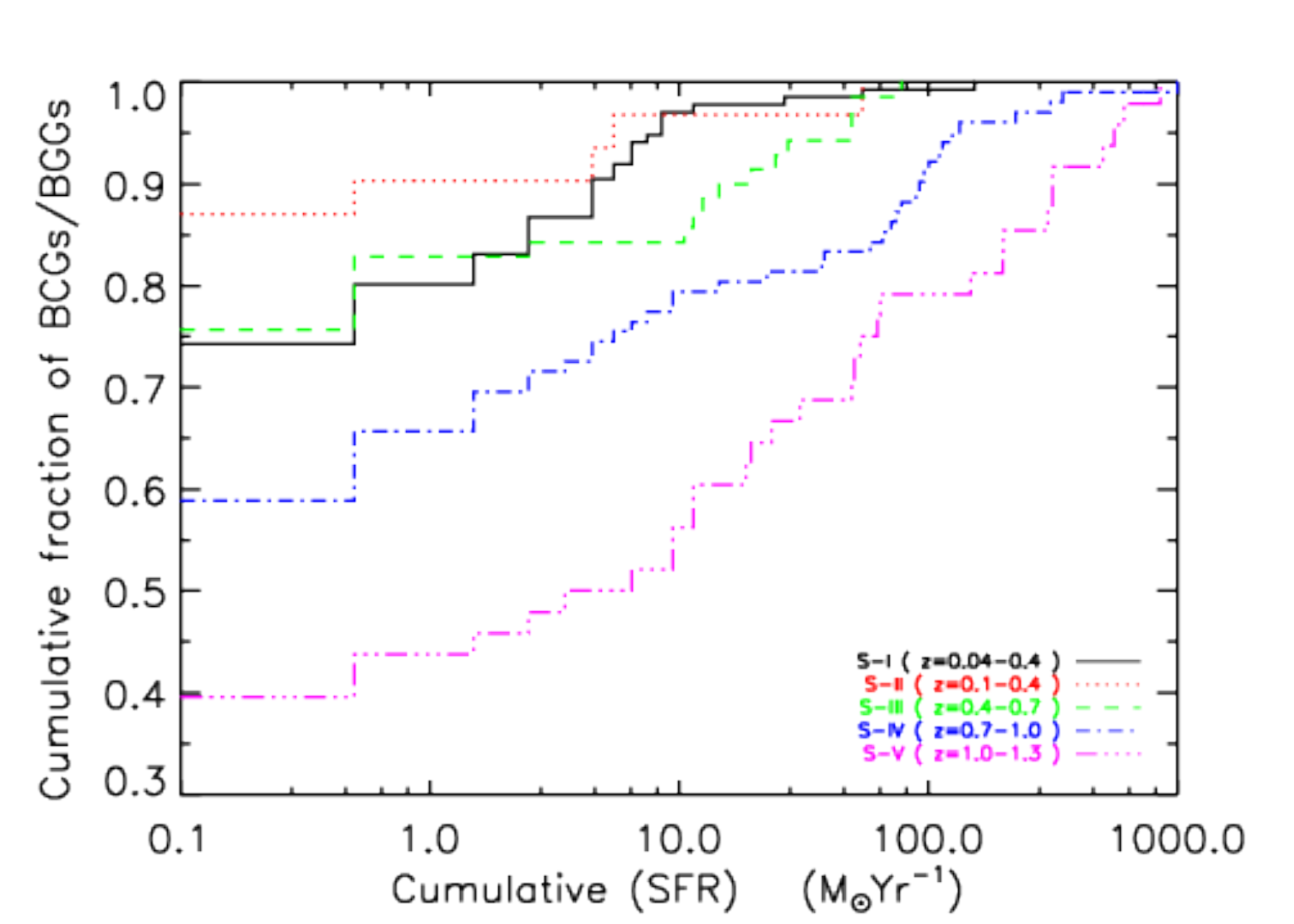}     
\caption[]{The specific star formation rate (sSFR)  distribution of the BGGs  in observations and  models ( G11 and DLB07) for S-I (top-left), S-II (top-right) , S-III (middle-left), S-IV (middle-right ),
and S-V ( bottom-left).  To adopt and illustrate the BGGs with SFR=0, the  SFR in both models and observations are shifted by $dSFR=+1^{-9} M_{\odot}yr^{-1}$. {\it (Bottom- right)} The cumulative  distribution of the  SFR ($M_{\odot}yr^{-1}$) of BGGs in observations. We
find that at least $ \sim20\pm3\% $ of the BGGs are star forming system with rates above $ >1 M_{\odot}yr^{1-} $. }
 \label{sfr-dis}
\end{figure*} 
\section{Star formation activity } \label{sfr}

\subsection{Distribution and evolution of the specific star formation rate (sSFR)}
A number of previous studies have shown that  BGGs are generally passive galaxies and a large fraction of these systems exhibiting  AGN activity, radio emission and no significant star formation at $ z<1 $. The newly formed stars in these galaxies contribute   $ \sim 1\% $ of the  total stellar mass of galaxy at late epochs $ z<2 $ \citep{Bildfell08,Edwards07,Kauffmann03,odea08,odea10,Liu12,Thom12}.
  
In this study , the star formation rate of the BGGs in the AEGIS filed has been estimated using the FAST code \citep{Kriek09}  taken from \cite{Wuyts11} and the star formation rate of the BGGs in the COSMOS and XMM-LSS fields have been obtained based on SED fitting \citep{Ilbert10}.  Both  G11 and DLB07  similarly assume stars from gas cooling in the disk  according to the empirical relation taken from \cite{Kennicutt98}. However, G11 model refined the SAM used in construction of DLB07 model and the new modifications of physical processes (e.g. the treatments of the gas cooling and disk size) in this  model lead  the SFR to evolve significantly smoother than that in DLB07, with less star formation activity driven from "star burst" in the bulk of the galaxies. For further information on the SFR estimate in models we refer to \S 3 in \citep{Guo11}. 

We present the distribution of the specific star formation rate of the BGGs in Fig. \ref{sfr-dis} and find the following.

For  S-I, the $log(sSFR/Gyr^{-1})$  spans  a wide range between $\sim-11$ to 1 with a peak around $log(sSFR/Gyr^{-1})\sim-3 $ (top-left panel).  A significant fraction ($\sim65\%$) of BGGs in models show no star formation activity. The discontinuity in the computed values for  the SFR in the SAM could well be a computational issue such that  sSFR less than $log(sSFR.Gyr^{-1})\le-4$ is assumed to be insignificant. We note that the SFR of BGGs in  models have been truncated at $log(SFR/M_{\odot}yr^{-1})sim-3$.  With this in mind, about $40\%$ of the observed sSFR would be below the SAMs sensitivity to the SFR list of $log(sSFR/Gyr^{-1})\le-4$ compared to the $\sim65\%$ in the models. 
  
For S-II, the distribution of the BGG sSFR in observations is broadly similar to that of S-I and tends to skew towards low sSFR. In models, fraction of highly star forming BGGs shows a slight increase relative to S-I (right-top panel).
 
For S-III, the distribution of the sSFR in observations  indicates that the star formation has been significantly higher in the past. We find that fraction of the  quiescent BGGs in models decreases by $\sim15\%$ compared to the low-z BGGs (S-I and S-II) which follows the observed trend. 

For S-IV and S-V, the sSFR distributions in observations and models further supports the increased star formation activities in BGGs at higher redshifts in both the observations and the models. 
      
We find that the sSFR distribution tends to skew towards low sSFRs and also tends to peak less than a normal distribution at all redshifts. In addition, we mention that the sSFR distribution tends to skew more towards low sSFR with increasing redshift.
 
Further, we study the cumulative distribution of the SFR  of BGGs in observations as shown in the bottom right panel of Fig. \ref{sfr-dis}. The vertical and horizontal axes represent  the cumulative fraction of the BGGs and the cumulative SFR in each subsample, respectively. The gap between these cumulative distributions clearly shows that the fraction of star forming BGGs increases with redshift. As a result,  we find that $ \sim 25\pm5$, $13\pm2$, $25\pm5$, $41\pm7 $, and $60\pm5 $ percent of the BGGs within S-I to S-V have total SFR above  $0.1 M _{\odot}yr^{-1}$ which can reach up to $1000M _{\odot}yr^{-1}$. By comparing the cumulative distribution of SFR between S-I (solid black histogram) and S-II (dotted red histogram), it appears that BGGs within low-mass groups are more star forming systems than the BGGs within  massive groups  in similar redshifts below $z<0.4$. 
 
 \begin{figure}
 \includegraphics[width=0.48\textwidth]{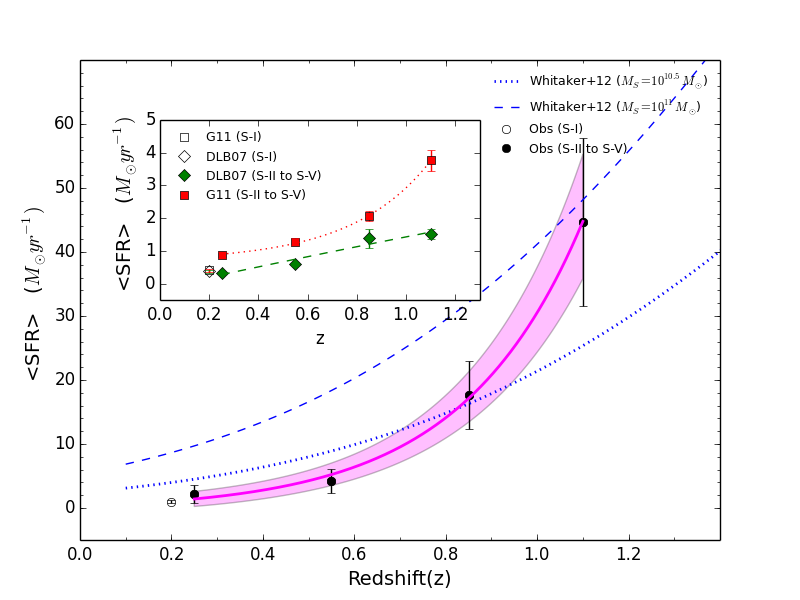}   
\caption[]{The redshift evolution of the average SFR of the BGGs in observations for  S-I (black open circle)  and for S-II to S-V  (black filled circles). The  magenta line and the
highlighted area represent the best-fit relation and its  68\% confidence intervals. The evolution in the SFR at two given masses  ($M_{S}=10^{10.5}$ and $10^{11} M_{\odot}$ ) in the SFR-mass sequence taken from \cite{Whitaker12} have been shown with dotted and dashed lines, respectively. The average SFR of the BGGs in observations is less than the SFR of the main sequence galaxies with similar masses in the filed and the BGG evolution is different from the main sequence evolution. The subplot shows the redshift evolution of the mean SFR of the BGGs in the G11 (red
diamonds and line) and DLB07(green squares and line) models, respectively.}
 \label{sfr-z}
\end{figure}

\subsection{Evolution of the average SFR of the BGGs}

To further explore the SFR evolution with redshift, we compute the average SFR of all BGGs in each subsample and show against their mean redshift in Fig. \ref{sfr-z}. We find that the mean SFR of BGGs within rich groups (S-II to S-V) steeply declines from $z=1.3$ to the present day, changing by roughly 1.98 dex $M_{\odot}yr^{-1}$. This evolution is significantly stronger than the decrease in the SFR of galaxies as whole for a given mass by a factor of $\sim30$ from $z\sim2$ to $z\sim0$ \citep[e.g.][]{Daddi07}. The best fit relation to the evolution of the mean SFR of BGGs in observations with reshift is approximated as follow:\\
$ <SFR>= (0.70\pm0.06)e^{(3.79\pm0.10)z}+(-0.39\pm 0.97) $.

The highlighted magenta area in Fig. \ref{sfr-z} corresponds to the 68\% confidence intervals. 

A systematic difference is seen between observations and model predictions, as explained in last section, thus we separately illustrate  the SFR evolution with redshift for SAMs in Fig. \ref{sfr-z}.  The best-fit relation to the mean SFR-z plane in the G11 model also follows an exponential trend ($ <SFR>= (0.098\pm0.001)e^{(3.13\pm0.1)z}+(-0.68\pm0.01)$). In contrast, the DLB07 predicts a linear evolution for the SFR of BGGs with redshift ($<SFR>= (1.53\pm0.28)z+(-0.10\pm0.21)$).  The mean SFR of the BGGs for S-I is shown with open symbols in both observations and models predictions.

In addition, to compare the evolution of the average SFR of the BGGs with the evolution in the SFR of the star forming galaxies, we use the SFR-stellar mass sequences (main sequence) driven using a sample of 28,701 galaxies selected from the NEWFIRM Medium-Band Survey  \citep[NMBS;][]{Whitaker11} in a wide redshift range ($0<z<2.5$) as presented in \cite{Whitaker12}. Since the majority of the BGGs in observations have a masses around $log(M_{S}/M_{\odot})\sim11$,  We use function (1) in  \cite{Whitaker12} and determine  the SFR evolution at $log(M_{S}/M_{\odot})=11$. In addition, we also estimate this evolution in the SFR-mass sequence  \citep{Whitaker12} at $log(M_{S}/M_{\odot})=10.5$.

 As a result, by comparing the evolution of the main sequence at $log(M_{S}/M_{\odot})\sim11$ (dashed blue line) and the evolution of the average  SFR of BGGs (solid magenta line), we find that the average SFR of BGGs is less than the SFR of the star forming galaxies in the field in similar stellar masses. This indicates that the BGG evolution is different from the main sequence evolution.

 \subsection{Correlation between star formation rate and stellar mass} \label{sfr-sm}
Fig. \ref{sfr-sm} highlights the relation between the SFR and the stellar mass of the BGGs.  The observed data is overlaid on the same from the SAM of G11 (red diamonds) and  DLB07 (green squares).  For clarity, the BGGs with photometric redshift  (black symbols) are distinguished from those with spectroscopic redshift (magenta symbols). We illustrate the SFR-stellar mass relation for the low-z sub-samples (S-I (circles) and S-II (triangles)), the intermediate-z sub-samples (S-III (circles) and S-IV (triangles)), and the high-z subsample (S-V) from top to bottom panels, respectively. 

We use the SFR-stellar mass sequences \citep{Whitaker12} to define star forming, star-burst and passive BGGs in our sample \citep[e.g.][]{Elbaz07,Erfanianfar14}. Galaxies spend most of their life on the main sequence and keep their star forming activity as normal star forming systems. As their star formation activities are quenched, they fall below the main sequence and become as passive/quiescent galaxies. In contrast,  as a galaxy experiences a merger or becomes in the close encounter with another galaxy, it may undergo an exceptionally high rate of star formation it moves above the MS and spend a short fraction of its life in this region. In this phase, this galaxy is identified as a star-burst galaxy. The dashed orange line in Fig \ref{sfr-sm} represents the position of the SFR-mass sequences  at $ 0.0<z<0.5 $ (upper panel), $  0.5<z<1.0  $ (middle panel), and  $1.0<z<1.5 $ (lower panel) taken from  \cite{Whitaker12}.

Therefore, we define a BGG as a star forming galaxy if it falls on the MS, i.e. between the two dashed blue lines which correspond to $ \pm1$ dex $M_{\odot}yr^{-1}$. We define a BGG as passive galaxy, if the SFR of this galaxy falls  below  the lower limit ($-1 dex$) from  MS for a given stellar mass. 

\begin{figure}
 \includegraphics[width=0.45\textwidth]{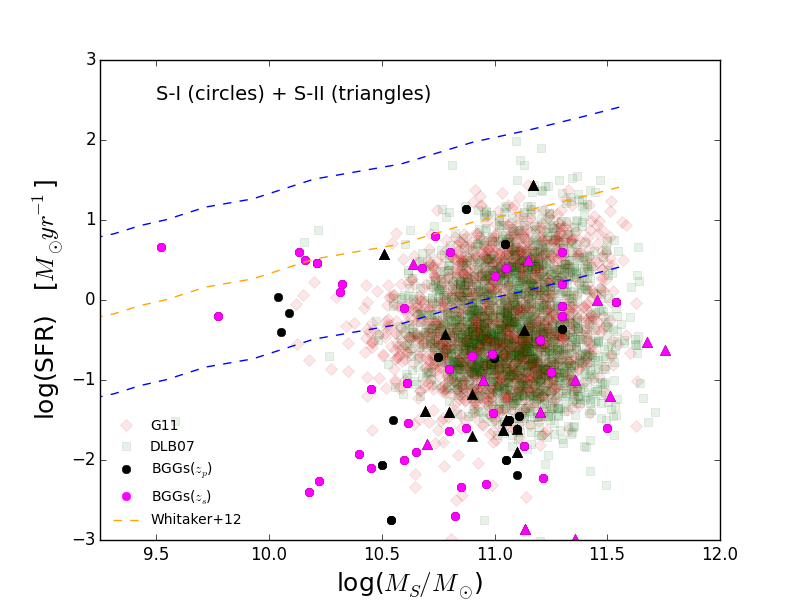}
  \includegraphics[width=0.45\textwidth]{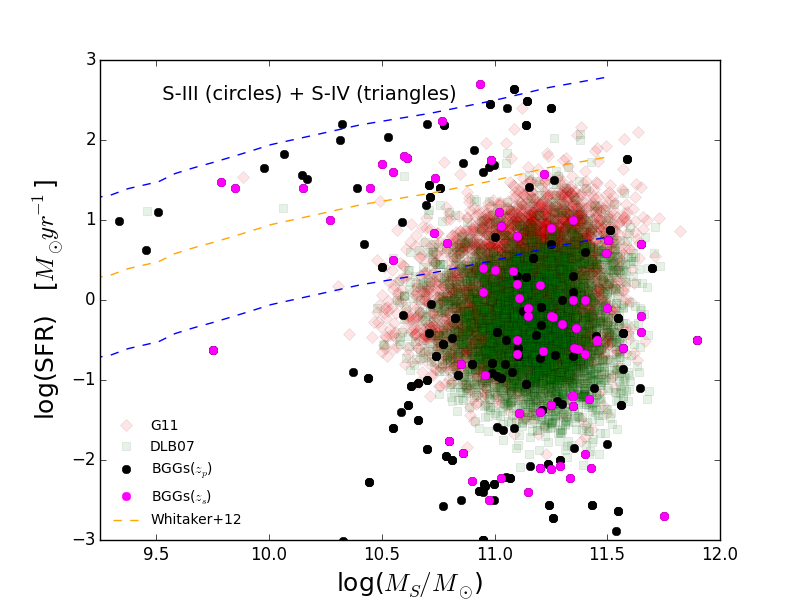}
    \includegraphics[width=0.45\textwidth]{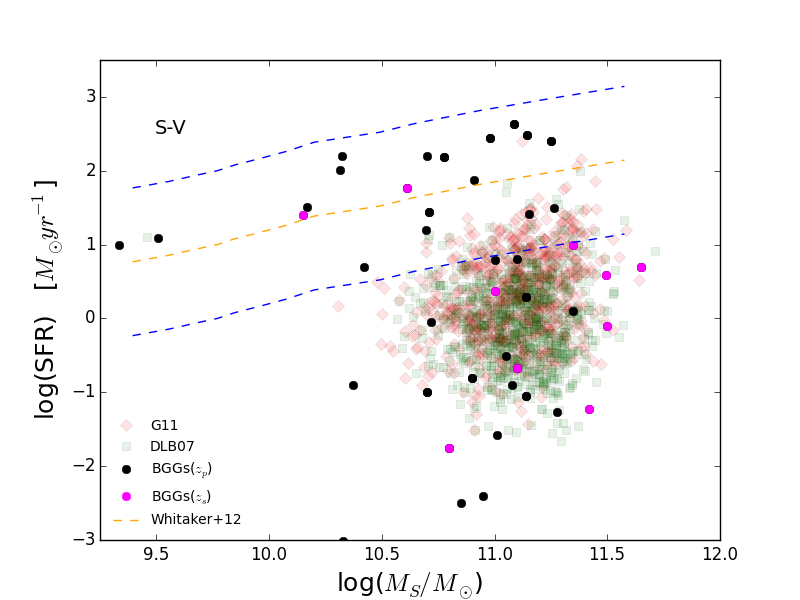}

\caption[]{The BGG SFR  versus their stellar mass  for observations (black and magenta symbols) overlying  on that from the SAM of G11 (red diamonds) and DLB07 (green squares). The magenta and black colors distinguish the BGGs with spectroscopic and photometric  redshifts, respectively. Upper to lower panels present the $SFR-M_{S}$ plane for S-I+ S-II, S-III+ S-IV,  and S-V, receptively. The dashed orange line represents the SFR mass sequence taken from \cite{Whitaker12}. The blue dashed lines represent $  \pm1 $ dex $M_{\odot}yr^{-1}$ limits from the SFR mass sequence. We select BGGs as a star forming galaxy if its SFR lies on the SFR mass sequence between two dashed blue lines, if the BGG SFR  falls below the lower level of this sequence it is classified as a passive galaxy.  }
 \label{sfr-sm}
 \end{figure}    
 
In Fig. \ref{sfr-sm},  we find that the fraction of  galaxies falling in the star forming region increases with redshift in observations,  while the trend is less clear in the models. We also find that the majority of the low-mass BGGs ($log(M_{S}/ h^{-1}M_{\odot})<10.5 $) lie within the SFR-mass sequences. We also find that galaxies with stellar mass of $log(M_{S}/ M_{\odot})=10.5-11$ exhibit a bimodal distribution between the low- and high mass galaxies. 

In addition,  the $sSFR-M_{S}$ plane for S-III to S-IV (middle panel)  shows the presence of  some highly star forming, star-burst, BGGs in observations close to ( or above) the  star formation main sequence.
\begin{figure}
\includegraphics[width=0.52\textwidth]{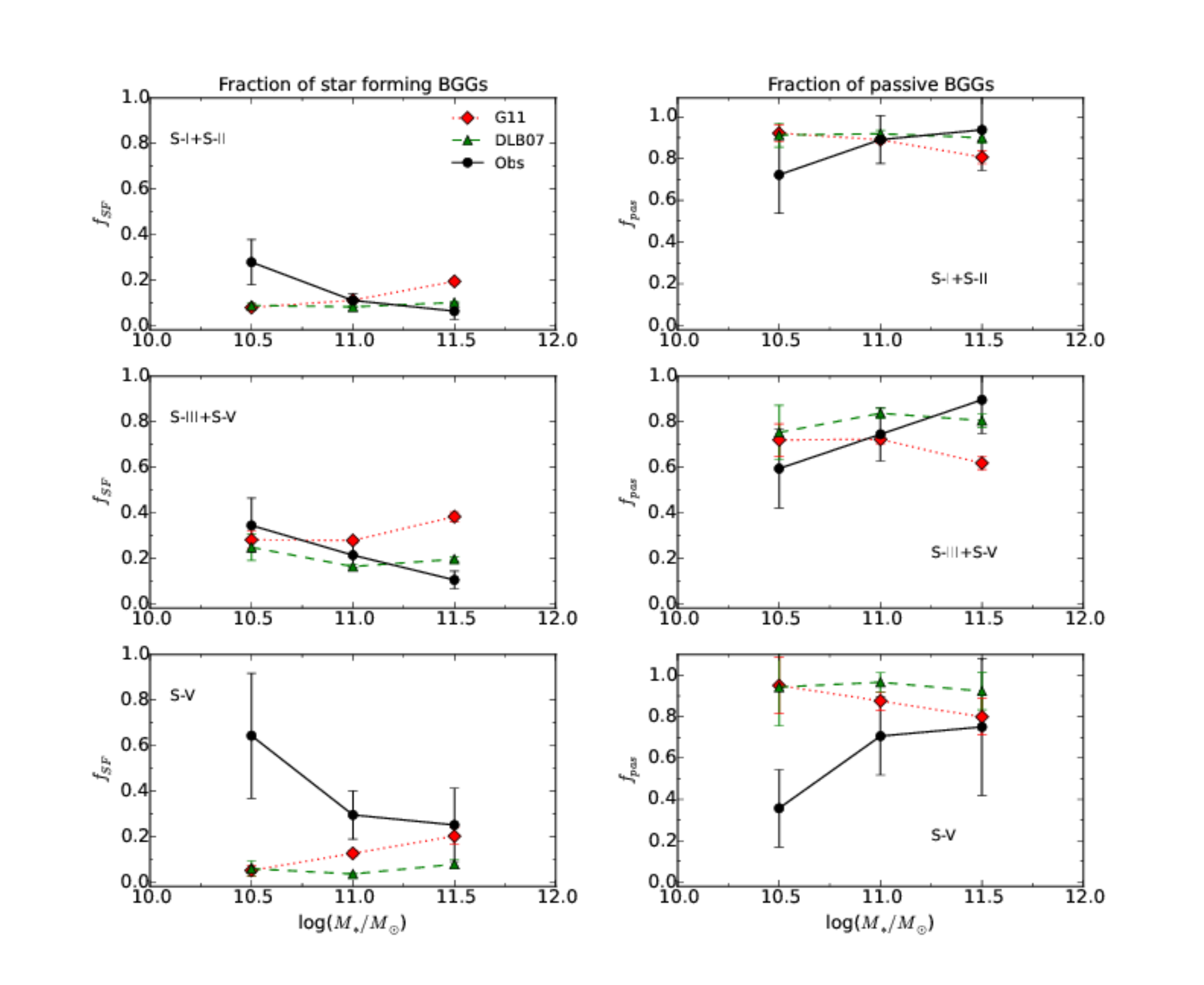}
\caption[]{The fraction of the star forming ($f_{SF}$) (left panels) and the passive ($f_{pas}$) (right panels) BGGs versus the stellar mass of BGGs for S-I+S-II, S-III+S-IV, and S-V. The solid black line, dotted red line, dashed green line illustrate the trends in observations and the G11 and DlB07 models, respectively.}
\label{pass-sf}
\end{figure}

A considerable number of the BGGs, in models,  show no star formation activity and they fall out of the SFR range which is adopted in Fig. \ref{sfr-sm}. Thus to have a better comparison between observations and model predictions  we bin stellar mass of BGGs and compute the number of  passive and star forming (SF) BGGs in each bin. The number of passive/SF BGGs in each stellar mass bin is normalised  to the total number of the BGGs in each subsample. The left and right panels of Fig. \ref{pass-sf} show the fraction of SF BGGs ($f_{SF}$) and the fraction of passive BGGs ($f_{pas}$) as a function of their stellar mass, respectively. In upper and middle panels in Fig. \ref{pass-sf}, we present results for the combined data of S-I+S-II and S-III+S-IV, respectively, since the trends for these combined subsamples were similar. 
As a result, for all subsamples, we find that the fraction of star forming BGGs decreases  with increasing stellar mass with a corresponding increase of passive BGGs. The passive fraction at fixed stellar mass seems to increase at lower redshift. Interestingly, G11 shows an increasing fraction of star forming with mass in contrast to the observation. The DLB07 prediction  agrees better than G11 with observations at $z<1$. The  trends in observations and models show no significant dependence to redshift.

Although we do not explicitly show in the paper, we find a similar trend, as seen in Fig. \ref{pass-sf}, in the fraction of star forming and passive BGGs as a function of the halo mass for both models and observations. 
   
\subsection{Specific star formation rate \textit{versus}  redshift and M$_{200}$}
\begin{figure*}
  \includegraphics[width=0.8\textwidth]{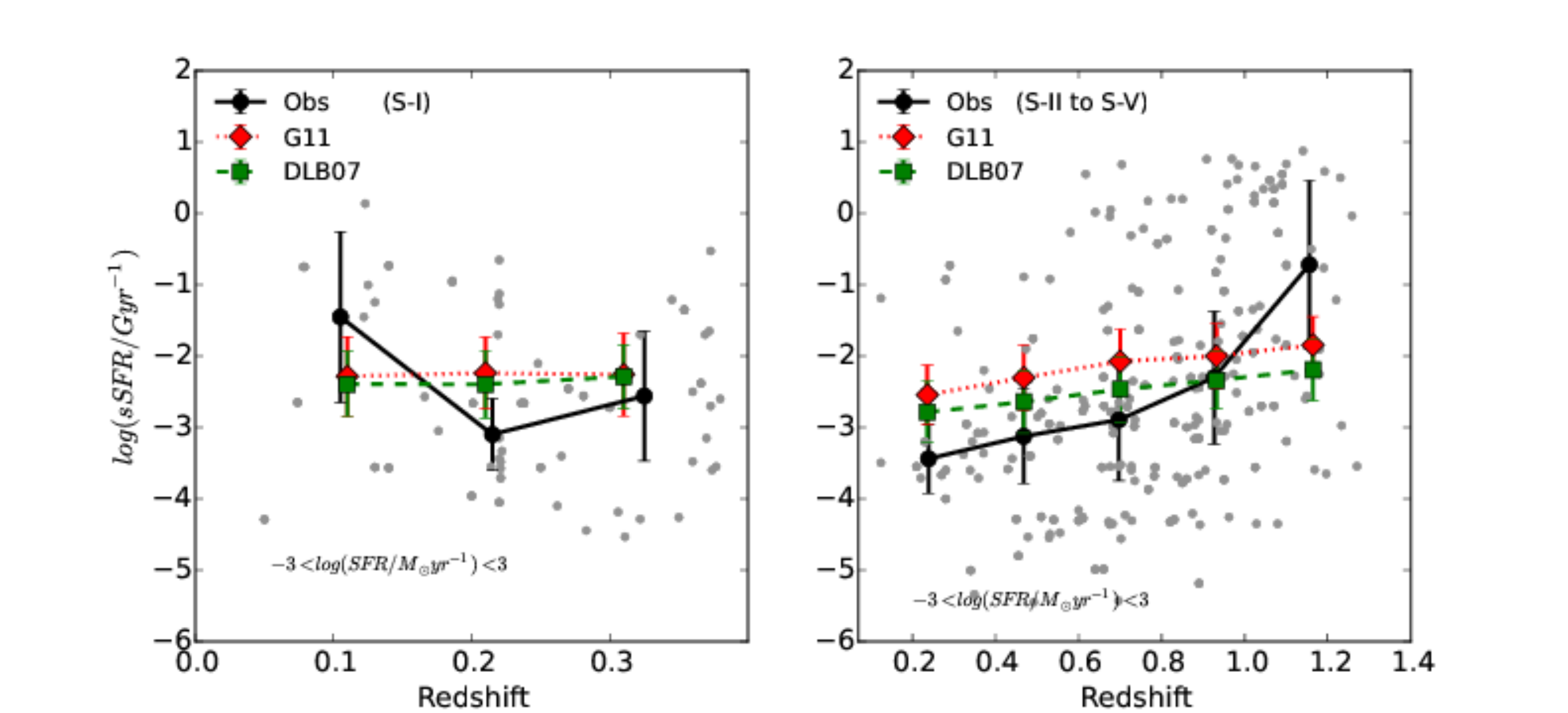}

 \caption[]{The $ SFR/M_{s}-z  $ relation for  S-I (left panel), and for S-II to S-V  (right panel). The solid, dashed, and dotted  lines show the median trends  for BGGs in observations and  in the SAMs of DLB07 and G11, respectively. The gray point shows the BGG sSFR versus redshift in observations. }
 \label{ssfr-z}
 \end{figure*}

Finally we study the evolution of the specific star formation rate and its relation with the halo mass. Fig. \ref{ssfr-z} shows the redshift dependency of the median  $ sSFR $ for S-I (left panel) and for S-II to S-V (right panel), respectively. The error bars in data points corresponds to the median absolute deviation in each redshift bin. In agreement with model predictions,  the  specific star formation rate of BGGs for S-I show no significant change with redshift . In  Fig. \ref{ssfr-z}, we also probe the $ log(sSFR)-z$ relation in the  SAMs. Models predict a flat  trend over $z<0.4$. The left panel in Fig. \ref{ssfr-z} shows that on the other hand, in S-II to S-V, the sSFR of BGGs  increases  mildly with increasing redshift between $z=0.1$ to $z=0.7$. SAMs predict  a slow evolution over $0.1<z<1.3$ which they may marginally agree with observations within errors.

\begin{figure*}

  \includegraphics[width=0.8\textwidth]{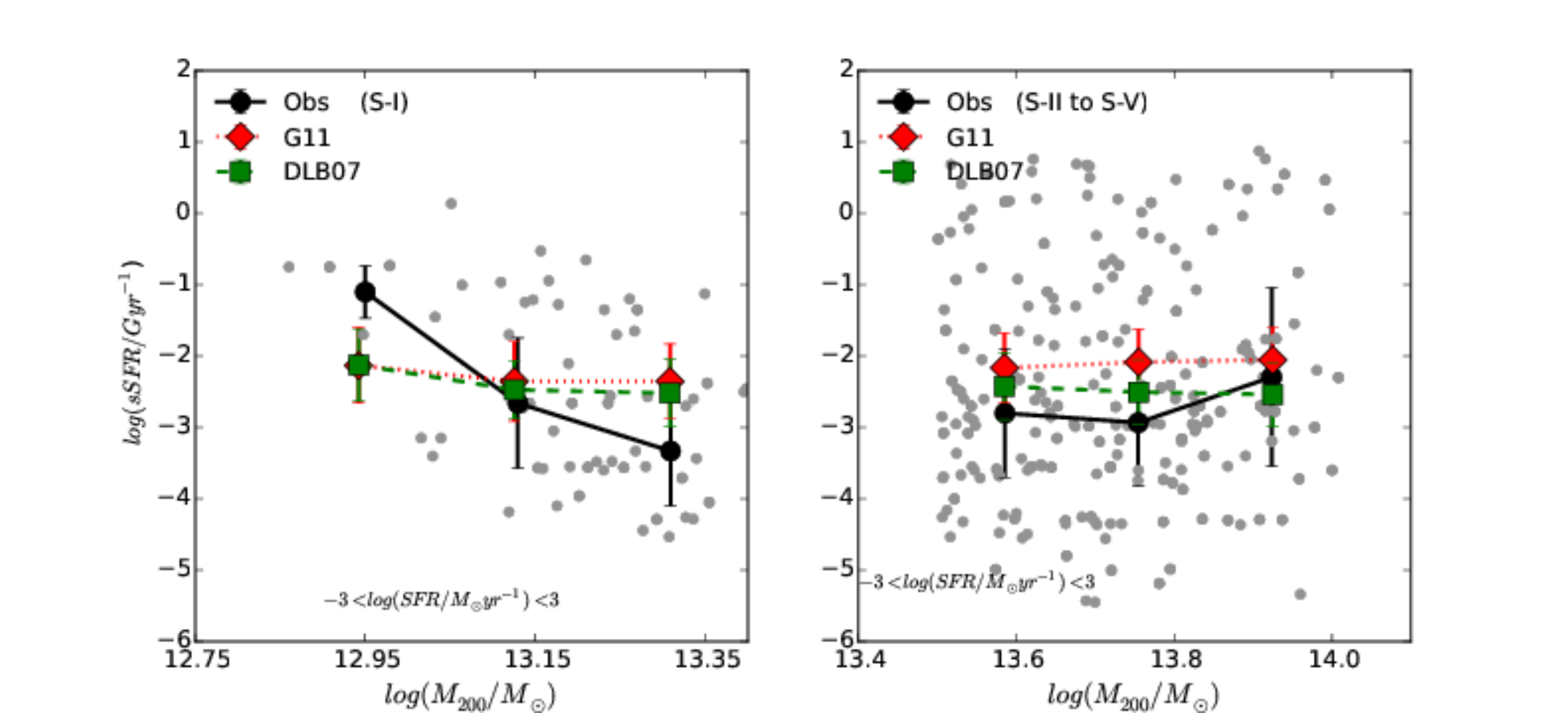}

 \caption[]{ The $ SFR/M_{s}-M_{200}$ relation  for S-I (left panel), and for  S-II to S-V  (right panel). The solid, dashed, and dotted  lines show the median trends  for BGGs in observations and  in the SAM of DLB07 and G11, respectively. The gray point shows the BGG sSFR versus redshift in observations. Only, for BGGs within S-I we find that the sSFR decreases as a function of halo mass.  }
 \label{ssfr-m200}
 \end{figure*}

In Fig.  \ref{ssfr-m200}, we show the  median sSFR dependency  of BGGs to the total mass of their host groups in observations (black points) and in the SAMs of G11 (red diamonds) and DLB07 (green squares), respectively. For S-I (left panel), the median sSFR rapidly decreases as a function of increasing halo mass (the best-fit linear relations corresponds to: $SFR/M_{s}=(-4.4\pm1.0)\times M_{200}+(3.8\pm0.8$). The observed trend is steeper than the model predictions. In agreement with  the SAM predictions, we find no relationship between sSFR and $M_{200}$  for S-II to S-V over $ 0.1<z<1.3 $, as shown in the right panel in Fig. \ref{ssfr-m200}. These results suggest that, in the group scale halos, sSFR of BGGs is a function of redshift with no halo mass dependence.
   
      
\section{Summary and conclusions}

We select the brightest group galaxies in a well defined  X-ray-selected sample of galaxy groups from the  COSMOS, XMM-LSS and AEGIS fields. This sample covers a redshift range of $0.04<z<1.3$ and contains about 407 galaxy groups with a mass ranging from 10$^{12.8}$ to 10$^{14}$M$_\odot$, the largest sample for such studies in the X-ray galaxy group regime. The sample allows us to study the distribution and evolution of the stellar mass and star formation rate of BGGs and the relation among the stellar mass, SFR, and halo mass. We compare our results with sample drawn from the Millennium simulations with three semi-analytic models (G11, DLB07 and B06). We summarise our results as follows:

(i) The shape of the stellar mass distribution of the BGGs evolves towards a normal distribution as the groups evolve. This evolution is quantified by the skewness and Kurtosis of the stellar mass distribution and their redshift dependence. At all redshifts, we find that the stellar mass distribution tends to skew towards  low masses. The observed  fraction  of the BGGs shows a strong deviation from the Gaussian  fit  at  $\approx 10^{10.2} M_{\odot}$,  which are generally star forming/young population. In contrast, the SAM predictions show no redshift evolution of the stellar mass distribution. Within the probed semi-analytic models, the shape of the stellar mass distribution predicted by \citet{Bower06} model is more consistent with the observations at $z<0.5$.

(ii) We show that the average stellar mass of the BGGs evolves with redshift by a factor of $ \sim2 $ from $ z=1.3 $ to the present day. At $ z<0.4 $, the mean stellar mass shows no significant evolution and  the significant growth of BGGs  occurs at $0.4<z<1.3$. The observed evolution is broadly consistent with the prediction of the $ \Lambda CDM $ model, and previous studies \citep[e.g.][]{Oliva14,Lin13,Lidman12}. Furthermore,  the SAMs of G11, DLB07 and B06 predict that BGGs grow in stellar mass by a factor of 1.55, 1.63, and 2.09 respectively. The rate  of the mean stellar mass  evolution with redshift  in B06 prediction has the best agreement with observations.
 
(iii) We find that the BGG stellar mass increases  with the halo mass, with a weak redshift dependency. In addition, the stellar mass of BGGs within low mass groups (S-I) correlates more strongly  with host halo mass compared to that of  BGGs within massive halos at similar redshift range ($z<0.4$). 

(iv) We show that the BGGs are not completely inactive or quenched systems as their SFR can reach $\sim 1000 M_{\odot} yr^{-1} $. At least $ \sim13\pm3 $ to $60\pm5$ percent of  BGGs in our low-z (S-II) and high-z (S-V) subsamples  are galaxies with star forming rate  above $ \sim1 M_{\odot} yr^{-1}$. We indicate that the average SFR of BGGs steeply increases  with  redshift in particular at $z>0.7$. The best fit to the data in observations is as follow: $ <SFR>= (0.70\pm0.06)e^{(3.79\pm0.10)z}+(-0.39\pm 0.97) $. We compare the evolution of the average SFR of BGGs with the evolution in the SFR  of the star forming galaxies (main sequence) using the SFR-mass sequence presented in \citep{Whitaker12}. We find that  evolution of the average SFR  of BGGs is different from the evolution of main sequence. In addition,  the SAMs underestimate the average SFR of BGGs in observations in similar stellar masses.
   
(v) We find that the fraction of star forming BGGs decreases  with increasing stellar mass with a corresponding increase of passive BGGs. In addition, the passive fraction at fixed stellar mass seems to increase at lower redshift. Interestingly, G11 shows an increasing fraction of star forming with mass in contrast to the observation, while fraction of  the passive BGGs increases with increasing stellar mass. The DLB07 prediction  agrees better than G11 with observations at $z<1$. We note that the relation between fraction of the star forming/passive  BGGs and the total mass of groups is similar to that is seen between these fractions and the stellar mass of BGGs.  

(vi) For BGGs in massive groups (S-II  to S-V) we find that sSFR slightly increases with increasing redshift at  $z<0.7$, and above this redshift, the trend of the sSFR evolution becomes steep. While the sSFR of BGGs for S-I shows no significant change with redshift. We also find that the sSFR of BGGs within low mass groups (S-I) decreases with increasing halo mass. However, in agreement with model predictions, the sSFR of BGGs within massive groups (S-II to S-V) shows no  dependence to the halo mass. 

We have been able to probe semi-analytic galaxy formation models, publicly available, using the observations of the stellar mass, star formation rate and their halo dependencies. We demonstrate and argue that these observations are highly useful for con-straining the models. 

\section{Acknowledgements}
This work has been supported by the grant of Finnish Academy of Science to the University of Helsinki, decision number 266918. The first author wishes to thank School of Astronomy, Institute for Research in Fundamental Sciences for their support of this research. We used the data of the Millennium simulation and the web application providing online access to them were constructed as the activities of the German Astrophysics Virtual Observatory.

\label{lastpage}

 \end{document}